# Electronic Reconstruction Enhanced Tunneling Conductance at Terrace Edges of Ultrathin Oxide Films

*Lingfei Wang[1,2], Rokyeon Kim[1,2], Yoonkoo Kim[3], Choong H. Kim[1,2], Sangwoon Hwang[1,2], Myung Rae Cho[1,2], Yeong Jae Shin[1,2], Saikat Das[1,2], Jeong Rae Kim[1,2], Sergei V. Kalinin[4], Miyoung Kim[3], Sang Mo Yang[4,5], and Tae Won Noh[1,2]*

[1]Center for Correlated Electron Systems, Institute for Basic Science (IBS), Seoul 08826, Republic of Korea;
[2]Department of Physics and Astronomy, Seoul National University, Seoul 08826, Republic of Korea
[3]Department of Materials Science and Engineering and Research Institute of Advanced Materials, Seoul National University, Seoul 08826, Republic of Korea
[4]Center for Nanophase Materials Sciences, Oak Ridge National Laboratory, Oak Ridge, Tennessee 37831, United States
[5]Department of Physics, Sookmyung Women's University, Seoul 04310, Republic of Korea

**Abstract**

Quantum mechanical tunneling of electrons across ultrathin insulating oxide barriers has been studied extensively for decades due to its great potential in electronic device applications. In the few-nanometer-thick epitaxial oxide films, atomic-scale structural imperfections, such as the ubiquitously existed one-unit-cell-high terrace edges, can dramatically affect the tunneling probability and device performance. However, the underlying physics has not been investigated adequately. Here, taking ultrathin $BaTiO_3$ films as a model system, we report an intrinsic tunneling conductance enhancement near the terrace edges. Scanning probe microscopy results demonstrate the existence of highly-conductive regions (tens of nanometers-wide) near the terrace edges. First-principles calculations suggest that the terrace edge geometry can trigger an electronic reconstruction, which reduces the effective tunneling barrier width locally. Furthermore, such tunneling conductance enhancement can be discovered in other transition-metal-oxides and controlled by surface termination engineering. The controllable electronic reconstruction could facilitate the implementation of oxide electronic devices and discovery of exotic low-dimensional quantum phases.

One of the fundamental fruitions in quantum mechanics is that electrons can tunnel across an insulating barrier with a thickness of a few-nanometers.[1–4] During the last two decades, due to the rapid advancements in heteroepitaxial growth techniques,[5–7] it has become feasible to reduce the oxide film thickness to nanometer scale, while retaining the original ferroic order. These advances enable oxide-based novel tunneling devices with superior data storage capabilities, including magnetic spin filters and ferroelectric (FE)/multiferroic tunnel junctions.[8–17]

A uniform one-unit-cell-high terrace structure is widely accepted as a prerequisite for realizing high-quality epitaxial oxide heterostructures.[5,18–20] However, given a well-defined terrace structure, the terrace edges can act as inevitable structural imperfections at the surface and interface. In ultrathin oxide films, the effect of these structural imperfections can be significant, inducing large changes in tunneling probability and device performance. In fact, conductance variations near the terrace edge has been observed in various ultrathin FE or dielectric oxides and considered as detrimental effects on the tunneling behavior.[11,13,21–23] Until now, most of these local conduction anomalies were vaguely attributed to extrinsic origins, such as defect-mediated leakage and topographic crosstalk.[21–24] However, the intrinsic effects of terrace edge geometry, including the broken bonds, missing atoms, and electrostatic boundary condition changes, have not been adequately investigated and understood.[25]

In this work, taking ultrathin $BaTiO_3$ (BTO) epitaxial film as a model system, we report an intrinsic tunneling conductance enhancement triggered by electronic reconstruction at terrace edges. Using spatially-resolved current-voltage (*I-V*) spectroscopy,[26] we can locally probe the quantum tunneling behavior near the terrace edge with a spatial resolution of ~10 nm. With the help of first-principles calculations and scanning transmission electron microscopy (STEM), we corroborate the local electronic reconstruction at the $TiO_2$-terminated BTO surface near the terrace edges. The resultant free electrons can reduce the effective tunneling barrier width ($t_B$) and facilitate the local tunneling conduction. We also demonstrated that such electronic reconstructions can be discovered in other transition-metal-oxides and controlled by surface termination engineering.

All oxide thin films studied herein were grown via pulsed laser deposition on atomically smooth $TiO_2$-terminated $SrTiO_3$(001) [STO(001)] substrates with a uniform terrace structure. We first deposited a 20-nm-thick $SrRuO_3$ (SRO) film as the bottom electrode and subsequently deposited the BTO films. We monitored the layer-by-layer growth of BTO using reflection high-energy electron diffraction and precisely controlled the film thickness ($t_{BTO}$) from 3 to 8 unit-cells (u.c.) (see Experimental Section, **Figure S1** and **S2** in the Supporting Information). To minimize the oxygen deficiency, we ex-situ annealed the films in the ambient oxygen flow at 600 °C for 1 h. The high epitaxial quality was confirmed via atomic force microscopy (AFM) and X-ray diffraction (XRD, **Figure S3** in the Supporting Information). The ferroelectricity of the ultrathin BTO films was established by the band-excitation piezoresponse switching spectroscopy results shown in **Figure S4-S6** in the Supporting Information.[27]

Using conductive atomic force microscopy (CAFM), we first checked the ferroelectric (FE) polarization-modulated tunneling current in the ultrathin BTO films. As schematically illustrated in **Figure 1a**, we wrote antiparallel FE domains in a preselected region by applying ±6 V DC tip biases (left panel), and then measured the local current through the domains (middle panel). Figure 1b-e shows AFM topographic images (top left panel) and corresponding CAFM images (bottom left panel) of BTO films with different $t_{BTO}$, ranging from 3 to 6 u.c.. All of the CAFM images show current contrasts over the antiparallel FE domains. The domains with upward polarization always show smaller current (OFF state) than those with downward polarization (ON state). The current difference is small for the 3

u.c. sample but becomes larger as $t_{BTO}$ increases. The observed tunneling electroresistance (TER) effect can be understood by the asymmetric screening of the polarization charges in the two different metallic electrodes (i.e., the CAFM tip and the SRO layer). The resultant asymmetric deformations of the electrostatic potential profiles inside the BTO barrier can enable a change in tunneling current via polarization reversal.[9]

In addition to the TER effect, we observed highly conductive regions within each FE domain. Taking the 3 u.c. BTO film as an example, the CAFM image (top left panel of Figure 1b) clearly shows curly stripe-like patterns with higher current, with the same shape as the terrace edges in the corresponding AFM image (bottom left panel of Figure 1b). We took the profiles of topography and current along the horizontal lines, with positions indicated by blue (red) arrows in the AFM (CAFM) images. The profiles (right panels in Figure 1b) clearly demonstrate that the highly conductive regions are located at the terrace edges. As shown in Figure 1c-e, this local conduction anomaly becomes weaker with increasing $t_{BTO}$ and eventually disappears in the 6 u.c. BTO sample.

The correlation between the TER effect and conductance enhancement near the terrace edges can be understood quantitatively. We first averaged the current peak values at the terrace edges ($I_{peak}$) and the background values within the terrace plateaus ($I_{background}$). The ratio $I_{peak}/I_{background}$ (Figure 1f) represents the conductance enhancement near terrace edges. Then we fabricated ferroelectric tunnel junctions (FTJ) based on the ultrathin BTO films (right panel of Figure 1a, see Experimental Section for details). After applying ± 6 V pulses to the top electrodes through the CAFM tip, we measured *I-V* curves at ON and OFF states ($I_{ON}$-*V* and $I_{OFF}$-*V*, see **Figure S7** in the Supporting Information). The ON/OFF ratio = $I_{ON}/I_{OFF}$ (Figure 1g) is widely used to evaluate the TER effect.[10–13] For $t_{BTO} \geq 6$ u.c., when the $I_{peak}/I_{background}$ value is close to 1.0, the ON/OFF ratio decreases exponentially with $t_{BTO}$. This trend is well fitted by the Simmons equation which is obtained by applying the WKB approximation to the direct tunneling (DT) model (dotted line in Figure 1g, see Section 5 in the Supporting Information for details).[28] As $t_{BTO}$ decreases below 6 u.c., the ON/OFF ratio starts to deviate from the Simmons tunneling model, and the $I_{peak}/I_{background}$ value increases sharply. For $t_{BTO}$ = 3 u.c., $I_{peak}/I_{background}$ reaches ~3.0 in both the ON and OFF states, while the ON/OFF ratio is close to 1. In this ultrathin limit, considerable current can flow through the terrace edge even in the OFF state, which reduces the difference between $I_{ON}$ and $I_{OFF}$ and smears the TER effect.

To obtain further insights into the local conduction mechanism, we performed spatially-resolved *I-V* spectroscopy on the 3 u.c. BTO film.[26] We first selected an area of 240 × 240 nm$^2$ over a terrace edge (**Figure 2a** and 2b), and subsequently conduct the *I-V* spectroscopy over a grid of 24 × 24 pixels (Figure 2c). Note that the size of one pixel is 10 × 10 nm$^2$. The averaged *I-V* curve over the entire grid (Figure 2d) shows a typical tunneling behavior without hysteresis. As shown in the current maps under various biases (Figure 2e-g), we can observe highly conductive regions along the terrace edge.

To obtain statistically significant information hidden in the local current maps, we averaged the current value as a function of horizontal distance *D* between the pixel and terrace edge. As shown in Figure 2h, we first marked the boundary between two terraces (TEB) using a solid line (blue) according to the topography. Then, by translating the TEB horizontally, we obtained a series of imaginary guidelines. Some of them are represented by the dashed lines in Figure 2h. Figure 2i shows the current values (under **a** tip bias of +0.3 V) averaged along these guidelines as a function of *D*. The averaged current reaches a maximum at the TEB, but the profile is highly asymmetric about the TEB. The current profile in the *D* < 0 range (the upper atomic plane) exhibits a gradual change. In the *D* > 0 range (the lower atomic plane), by contrast, the current profile shows a steep decrease within a 10 nm wide pixel. This result signifies that the high conductance region can spread laterally from the TEB towards the upper atomic plane over several tens of nanometers.

According to the lateral current profile, we can define two regions with distinct spatial current distributions. In the lower atomic plane near the TEB ($D > 0$), the local current value remains small and nearly constant. We define this area as the terrace plateau region (TP region, marked by the blue area in Figure 2h). Note that this region is far from the next terrace edge on the right side (Figure 2a). In the upper atomic plane near the TEB ($D < 0$), the local current value decreases gradually from the maximum at $D = 0$. Hence, we define a four pixel-wide stripe along the TEB, with relatively high current, as the terrace edge region (TE region, marked by the orange stripe in Figure 2h). Figure 2j shows the averaged *I-V* curves over the TE ($I_{TE}$-*V*) and TP ($I_{TP}$-*V*) regions. Both the $I_{TE}$-*V* and $I_{TP}$-*V* curves are nonlinear and nearly symmetric, which is a typical quantum mechanical tunneling behavior.[17]

To clarify the underlying mechanism of high conductance observed at the terrace edges, we performed *I-V* curve fitting based on theoretical tunneling models. Conduction through ferroelectric/dielectric thin films can be mainly categorized into three mechanisms: DT, Fowler-Nordheim tunneling (FNT), and thermionic emission (TI).[17,29] For our 3 u.c. thick BTO film, DT conduction is prominent within the small bias range ($|V| \leq 0.7$ V), while FNT should dominate the conduction in the large bias range ($|V| > 0.7$ V) (see Section 5 in the Supporting Information for detailed discussions).[29] Considering the small FE polarization of 3 u.c. BTO sample (Figure S5 in the Supporting Information), we approximated the BTO layer as a rectangular-shaped potential barrier, and then fitted the *I-V* curves using the Simmons equation.[10,28] There are three variable fitting parameters in the Simmons equation: barrier width $t_B$, average barrier height $\varphi_B$, and effective electron mass $m_e^*$. Since the optimum $m_e^*$ is very close to the free electron mass for all the fitting results (see Table S1 in the Supporting Information), we will not discuss this parameter here for simplicity. As shown in Figure 2j, the experimental $I_{TP}$-*V* curve can be well fitted by the Simmons equation with a barrier width $t_B = t_{BTO} = 1.2$ nm (3 u.c.) and a reasonable barrier height $\varphi_B = 0.375$ eV.[10,23] Then we tried two methods to fit the $I_{TE}$-$V$ curves. (1) We initially assumed that the large current can be interpreted as additional leakage channels with simple Ohmic $I$-$V$ behavior,[30] connected in parallel with the tunneling conduction channel. However, this method leads to poor fitting and in particular a large deviation under small bias. (2) We then assumed that the terrace edge conduction is dominated by tunneling and the large current originates from changes in the tunneling barrier. As shown in Figure 2j, optimum fitting can be only achieved by reducing $t_B$ from 1.20 to 0.98 nm (by ~ 0.5 u.c.) and keeping $\varphi_B$ almost unchanged (from 0.375 eV to 0.370 eV). More detailed fitting results and analyses can be found in the Supporting Information, Section 5. Accordingly, we can assert that the higher conductance near the terrace edges originates from local changes in the tunneling barrier width rather than additional leakage channels.

After ruling out extrinsic effects (Section 6 and **Figure S9-S12** in the Supporting Information), we argue that the local tunneling barrier variation could originate from intrinsic carrier doping via electronic reconstructions. In transition-metal-oxides with multi-valence states (such as $Ti^{3+}$ and $Ti^{4+}$), electronic reconstructions can easily occur when the electrostatic boundary condition changes at surface or interface. One prominent example is the formation of 2-dimensional (2D) electron gas in STO-based systems. Electronic reconstructions and spatially confined carriers can be induced by the polarity discontinuity (STO/LaAlO$_3$),[31] interfacial redox reactions (amorphous oxide/STO),[32] UV light exposure (bare STO surface)[33]. In the following parts, we will show that the terrace edge geometry can trigger an electronic reconstruction at the BTO surface, resulting in the enhanced local tunneling conductance.

The 2D electronic reconstructions in oxide heterointerfaces or surfaces are highly dependent on the atomic stacking sequences.[31] Based on this consideration, we first investigated the surface termination of our ultrathin BTO films using STEM. According to the cross-sectional high-angle annular dark field (HAADF) images and atomically-resolved mappings of energy dispersive spectroscopy (**Figure S13** and **S14** in the Supporting Information), the BTO surface is uniformly terminated by the TiO$_2$ atomic plane. This observation is consistent with our recent results: the BaO-terminated surface is thermodynamically unstable under our growth condition, thus leading to the uniform TiO$_2$-terminated surface.[34] On this basis, the terrace edges are expected to be uniformly TiO$_2$-terminated as well.

In the TiO$_2$-terminated BTO films, the occurrence of the electronic reconstruction at the terrace edges can be easily explained using a simple ionic model, which considers the ionic charges and covalence states only. As displayed schematically in **Figure 3a**, we consider a stoichiometric BTO lattice with one-unit-cell-high terraces at the TiO$_2$-terminated surface. In the stoichiometric bulk BTO unit-cell (marked by box #1), both Ba and O have stable single valence states; thus, Ti always has a valence of +4. The Ti valence state does not change in the TiO$_2$-terminated surface unit-cell (marked by box #2) because both the SrO and TiO$_2$ atomic planes are charge neutral. By contrast, for the unit-cell located at the terrace edge (marked by box #3), two out of six O$^{2-}$ and six out of eight Ba$^{2+}$ are missing. This configuration leads to an uncompensated charge of +0.5 $e$. The most energetically favorable way for this charge compensation should be the electronic reconstruction of Ti cation at the terrace edge. In this conceptual ionic picture, 0.5 electrons per unit-cell could be transferred to the Ti 3$d$ band, and the valence state of Ti cation at the terrace edge could be reduced to +3.5.

This electronic reconstruction picture was corroborated by first-principles density function theory (DFT) calculations. We constructed a supercell by stacking a 6 × 1 × 6 u.c. BTO lattice and a 20 Å vacuum layer along [001] direction (Figure 3a). We introduced the terraces at the two back-to-back TiO$_2$-terminated surfaces by altering the in-plane supercell vector (see

Experimental Section for details).[25] The terrace width ($W$) was set as 6 u.c. to reduce the calculation cost. One electron was manually added into the supercell to provide the charge required for electronic reconstruction (0.5 $e$ per terrace edge unit-cell). The density of states (DOS) profiles projected along the TiO$_2$ planes are shown in the left panel of Figure 3b. In the four inner TiO$_2$ planes, the DOS profiles show typical band insulator characteristics. They are quite close to those calculated from the pristine supercell without terrace structure. By contrast, for the two terraced surface TiO$_2$ planes, the DOS of Ti 3d band near the Fermi level ($E_F$) increase substantially. The free electron density of the terraced surface layer ($\sigma$) is ~ 0.4 $e$ per $W \times 1$ surface u.c. (Figure 3b, right panel), which is close to the ideal value (0.5 $e$ per $W \times 1$ surface u.c.) from the simple ionic model.

The lateral current profile in Figure 2i shows that the tunneling conductance decreases gradually from the terrace edges to the terrace plateaus. The calculated local carrier densities at the surface Ti sites (**Figure S16** in the Supporting Information) also follow the same trend, although the decay is much faster due to the narrow $W = 6$ u.c. for calculation. Similar to the STO surface/interface, the BTO film surface might have a ladder-like 3$d$ band structure.[33,35] Since the carrier density is relatively low (0.5 $e$ per $W \times 1$ surface u.c., $W$ ~ 1000 u.c. for real BTO films), these electrons will predominately fill the lowest in-plane $d_{xy}$ states.[35] As a result, most of the free electrons are confined within the surface TiO$_2$ layer but spread laterally from the terrace edge unit-cell towards the upper TiO$_2$ atomic plane. According to Figure 2i, the spread width of free electrons on the real BTO surface can be as wide as tens of nanometers. Because this width is still far less than the longitudinal length of the terrace (over tens of micrometers), we suggest that this electronic reconstruction could be quasi-one-dimensional (1D) in nature.

The quantitative picture for explaining the tunneling conductance enhancement at terrace edges is shown schematically in Figure 3c. The free electrons originating from the electronic reconstruction are mainly confined within the BTO surface layer but spread laterally over the TE region. The average carrier density in the surface BTO layer (0.5 u.c.-thick) of the TE region is estimated to be ~ $1.6 \times 10^{20}$ cm$^{-3}$ (Section 8 in the Supporting Information), which is high enough to make BTO metallic.[36] As a result, the effective $t_B$ at the TE region is reduced by ~ 0.5 u.c.. Given an ultrathin film ($t_{BTO}$ < 5 u.c.), the reduced $t_B$ can substantially increase the tunneling conduction in the TE region and smear the average TER effect of an FTJ device. On the other hand, for $t_{BTO} \geq 6$ u.c., the tunneling current contribution from terrace edges becomes too small to be distinguished by CAFM and thus has little effect on the corresponding FTJ performance. This trend can be confirmed by both the CAFM results (Figure 1) and *I-V* spectroscopy on thicker BTO samples (**Figure S17** and **S18** in the Supporting Information).

Since the terrace structure is inevitable for most transition-metal-oxide epitaxial systems, the electronic reconstruction along terrace edges could also occur ubiquitously. Here we choose ultrathin STO film to verify this idea because it also allows the Ti valence to change in the same manner. Moreover, we can easily realize both uniformly TiO$_2$- and SrO-terminated surfaces in the ultrathin STO films (Figure S2 in the Supporting Information), which enables us to control the occurrence of the tunneling conductance enhancement at the terrace edges.

We first deposited the STO film homoepitaxially on an Nb-doped (0.5 wt %) STO(001) (NSTO) substrate with uniform TiO$_2$ termination. The layer-by-layer growth mode enables a consistent TiO$_2$ surface termination of STO film (**Figure 4a**). As in the case of TiO$_2$-terminated BTO films, we clearly observed substantially enhanced tunneling conductance at terrace edges (Figure 4b). To obtain a purely SrO-terminated STO surface, we first deposited a 10 nm SRO buffer layer, which leads to a SrO surface termination due to the high volatility of the RuO$_2$ layer during growth.[37,38] Since the SrO-terminated surface has good thermostability during growth, it can be preserved in the subsequently deposited STO film.[38]

This termination sequence is further confirmed by HAADF images in **Figure S15** in the Supporting Information. In this case, as shown in Figure 4c, the six-fold oxygen coordination of Ti is preserved, and hence the electronic reconstruction is prohibited. As a result, the CAFM image (Figure 4d) exhibits a very weak current contrast, possibly from topographic crosstalk. These results not only confirm the validity of our electronic reconstruction picture but also imply that proper surface termination engineering could prevent the detrimental effects of terrace edge conductance on oxide device performance.

In summary, we investigated intrinsic tunneling conductance enhancement at the terrace edges of ultrathin BTO films. Such enhanced local tunneling conductance originates from quasi-1D electronic reconstruction triggered by terrace edge geometry. Our results also highlight that the highly-conductive channels at terrace edges are the origin of performance degradations in ultrathin oxide-based tunneling devices, a fact that is often overlooked.

Notably, the terrace edge-triggered local electronic reconstruction could potentially exist in other transition-metal-oxide heterostructures. In these correlated electron systems, the valence states of transition metal and the electronic transport properties are intrinsically coupled. In addition to the aforementioned titanates, one possible example could be manganites, such as $(La,Sr)MnO_3$.[39] The electronic reconstruction at terrace edges may reduce the valence of Mn cation and facilitate a more insulating phase with polaron-mediated transport. Moreover, the local electronic reconstruction and its lateral dimension can be easily controlled by engineering the surface termination and varying the substrate miscut angle. Such tunability may provide a new ground for developing oxide-based quasi-1D electron system, which holds great potential in oxide electronics implementation and searching for exotic low-dimensional quantum phases.[40,41]

**Experimental Section**

*Oxide film growth and characterization:* Oxide films were fabricated using a pulsed-laser deposition system with a KrF excimer laser. Before deposition, STO(001) and NSTO(001) substrates with low-miscut-angle (0.05~0.1°) were etched using buffered hydrofluoric acid and annealed at ambient to create atomically smooth $TiO_2$-terminated surfaces with one-unit-cell-high terraces. The typical terrace width *W* is ~400 nm, namely, 1000 u.c.. The temperature of the substrate was maintained at 700 °C during deposition. The laser fluence and repetition rate were set at ~1.5 J/cm$^2$ and 2 Hz, respectively. The SRO bottom electrode was grown at an oxygen pressure ($P_{O2}$) of 100 mTorr. Ultrathin BTO (or STO) layers were subsequently deposited at $P_{O2}$ = 5 mTorr. The time-dependent intensity of the reflection high-energy electron diffraction (RHEED) patterns (Figure S1 in the Supporting Information) indicates that the SRO film was grown in a step flow mode, whereas the BTO and STO were grown in a 2D layer-by-layer mode. After deposition, the heterostructures were ex-situ annealed at 600 °C in an ambient oxygen flow for 1 h to minimize the oxygen deficiency. The film structure was characterized using a STEM (JEM-ARM200F, JEOL) and high-resolution XRD ($\lambda$ = 0.154056 nm, D8 DISCOVER, Bruker).

*Device fabrication:* A conventional electron-beam (e-beam) lithography resist (7% of 950k polymethyl methacrylate (PMMA) dissolved in Anisol) was spin-coated onto the BTO film surface and then baked at 180 °C. The resist film thickness was controlled at ~400 nm. E-beam exposure was performed using a modified scanning electron microscope with a nanometer pattern generation system (NPGS, JC Nabity Lithography Systems, MT, USA). Then the patterns were developed in a methyl isobutyl ketone (MIBK) solution diluted with isopropyl alcohol (IPA) (room temperature, MIBK:IPA = 1:3. Using this PMMA pattern as a lift-off mask, we sequentially deposited Ti (5 nm) and Au (25 nm) films onto the film surface using a conventional e-beam evaporator. The chamber pressure was maintained below $10^{-6}$

Torr to prevent oxidation of the metallic film. After deposition, the samples were immersed in acetone for lift-off. Following these procedures undertaken in a cleanroom, electrode patterns down to 500 nm in diameter, with a high yield over 90%, were obtained. We also performed additional AFM and CAFM measurements to confirm that the lift-off procedures do not cause detectable contamination and formation of passive layers on the BTO film surface (see Section 10 in the Supporting Information for details).

*Scanning probe microscopy (SPM) measurements:* All the SPM studies were performed at room temperature using a commercial scanning probe microscope (Cypher, MFP-3D, Asylum Research) equipped with a National Instruments data acquisition card, which is controlled by a custom-written LabVIEW and MATLAB programs. During the CAFM and *I-V* spectroscopy measurements, the voltage bias was applied from a conductive diamond-coated tip (boron-doped conductive diamond-coated tip, Bruker, Billerica, MA, USA). We fixed the contact forces between tip and sample at ~1000 nN (~500 nN) for CAFM (*I-V* spectroscopy) measurement, in order to minimize the current variation originating from poor tip/sample contact. The current signals were detected using a commercial current amplifier (Femto, DLPCA-200). To ensure a sufficiently large CAFM signal well above the noise background (10 pA), as labeled in Figure 1b-e, we varied the tips bias during current mapping ($V_{read}$) for samples with different $t_{BTO}$. Details of band-excitation piezoresponse switching spectroscopy can be found in the Supporting Information, Section 2.

*First-principles calculations:* The first-principles DFT calculations were performed using the Vienna *ab*-initio simulation package (VASP).[42] The supercell for calculation was constructed from a 6 × 1 × 6 BTO lattice and a 20 Å vacuum layer stacked along the [001] direction. The in-plane supercell vector was altered from $a_s = a$ to $a_s = 6a + c$ (*a* and *c* are the in-plane and out-of-plane lattice vectors of the BTO unit-cell, respectively, and $a_s$ is the in-plane vector of the supercell).[25] This lattice modification can introduce two back-to-back sets of terraces (*W*

= 6 u.c.) along the [010] direction while maintaining the stoichiometry. One electron was manually added into the supercell to mimic the self-doping induced by electronic reconstruction (0.5 $e$ per $W \times 1$ surface u.c.). The ferroelectricity of BTO was neglected for simplicity. Note that including FE polarization does not change the calculation results qualitatively. The projector-augmented-wave method was employed, together with the exchange correlation functional of the generalized gradient approximation in the Perdew-Burke-Ernzerhof scheme.[43] The self-consistent total energy was evaluated with a $2 \times 12 \times 1$ k-point mesh, and the cutoff energy for the plane-wave basis set was 500 eV.

**Acknowledgements**

This work is supported by the Institute for Basic Science in Korea (Grant No. IBS-R009-D1). This research was also supported (S.M.Y.) by the SookmyungWomen's University Research Grants (1-1703-2018). A portion of this research was conducted at the Center for Nanophase Materials Sciences, which is a DOE Office of Science User Facility. We would like to thank Prof. Je-Gen Park and Mr. Sung Min Lee for the support in SPM measurements. We also acknowledge the invaluable comments and suggestions of Dr. Bongju Kim, Prof. Seo Hyoung Chang, Prof. Tae Heon Kim, Prof. Jong-Gul Yoon and Prof. Jin-Seok Chung.

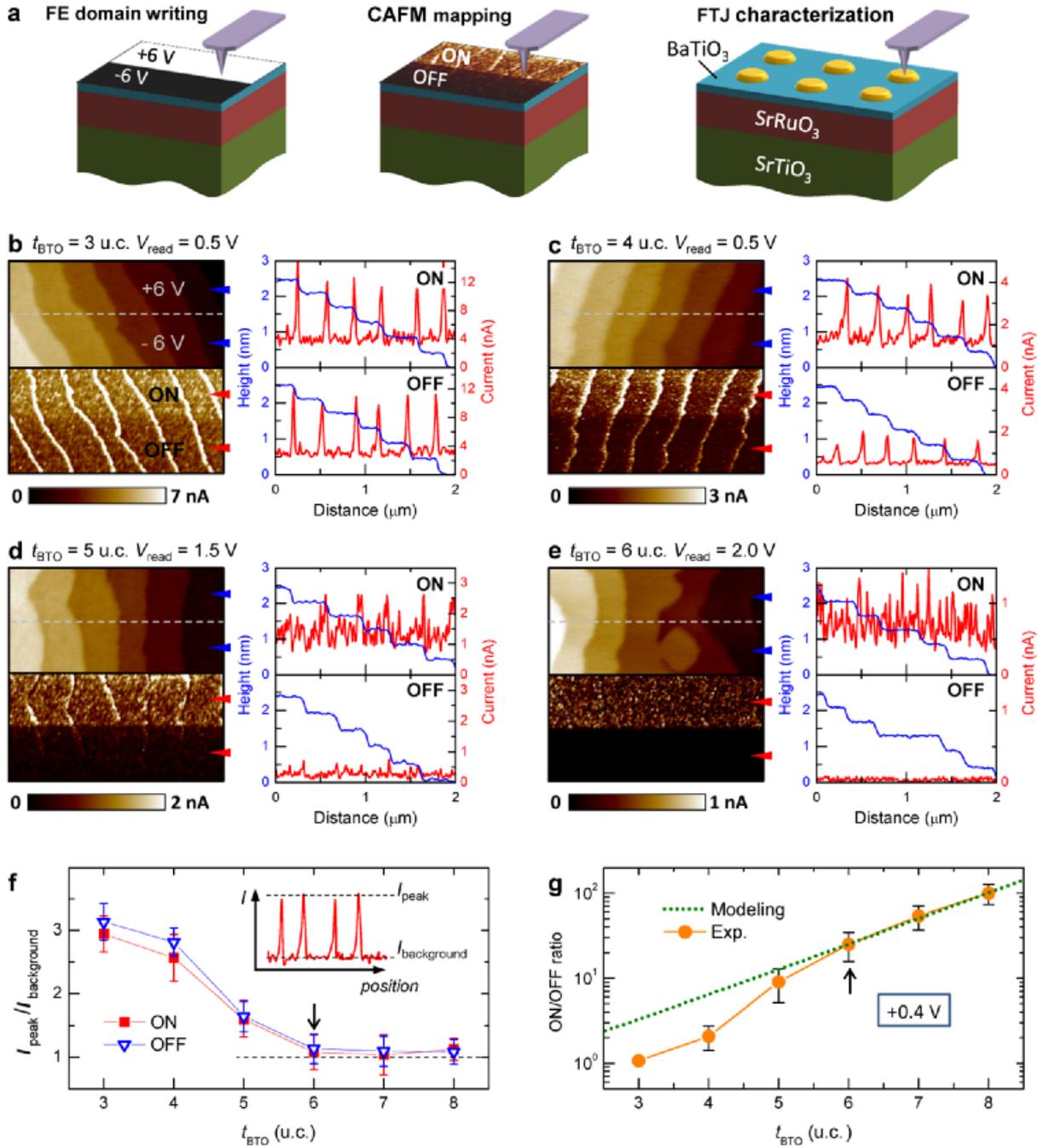

**Figure 1.** a) Schematic illustrations of conductive atomic force microscopy (CAFM) measurements and ferroelectric tunnel junction (FTJ) characterization. b–e) Atomic force microscopy (AFM) topographic images (top left panel) and CAFM images (bottom left panel) of the BaTiO$_3$/ SrRuO$_3$/ SrTiO$_3$ (BTO/SRO/STO(001)) films with various film thickness ($t_{BTO}$). The scanning area is 1 × 2 µm$^2$. The tip bias used for CAFM current mapping ($V_{read}$) was labeled for each sample. The boundaries of these antiparallel ferroelectric (FE) domains are marked by dashed (gray) lines. AFM (CAFM) horizontal line profiles along the blue (red) triangles are shown in the right panels. f) $t_{BTO}$-dependent $I_{peak}/I_{background}$ derived from the current line profiles shown in (b-e). The inset illustrates the method for defining $I_{peak}$ and $I_{background}$. The error bars represent the standard deviations of the $I_{peak}/I_{background}$ value distributions derived from the current line profiles. g) $t_{BTO}$-dependent ON/OFF ratios under a tip bias of +0.4 V averaged over 50 FTJ. The error bars correspond to the standard deviation of ON/OFF ratios from these devices. The dotted (green) line indicates the $t_{BTO}$-dependent

ON/OFF ratio predicted by the Simmons tunneling model. As $t_{BTO}$ decreases below 5 u.c. [marked by solid arrows in (f) and (g)], the $I_{peak}/I_{background}$ value starts to increase above 1.0, and the experimental ON/OFF ratio starts to show obvious degradation compared to the theoretically predicted value.

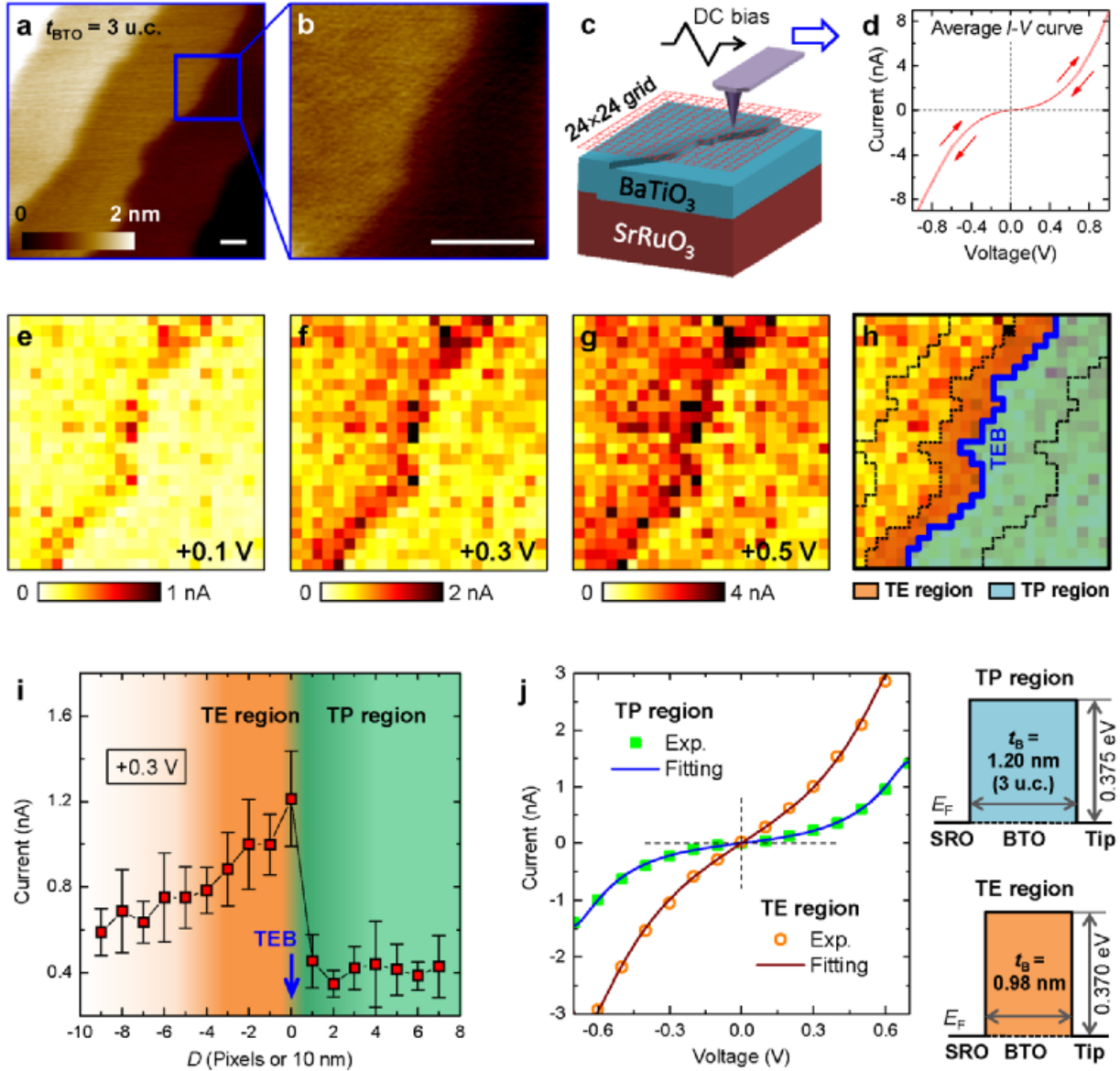

**Figure 2.** a,b) AFM topographic images of the 3 u.c. BTO film. All scale bars correspond to 100 nm. Prior to the current-voltage (*I-V*) spectroscopy measurements, we first measured the topography from an area of 1 × 1 µm² [shown in (a)] and then selected an area of 240 × 240 nm² [shown in (b)] over a terrace edge. c) Schematic illustration of the *I-V* spectroscopy. The scanned area in (b) was divided by a grid of 24 × 24 pixels and *I-V* measurements were performed in each pixel through the CAFM tip. d) *I-V* curve averaged over all 24 × 24 pixels. e-g) Spatial maps of local current under different tip biases. Similar to the CAFM images, all of the maps clearly show higher current at terrace edges. h) Schematic illustration for defining the boundary between two terraces (TEB, solid line, blue), terrace edge region (TE region, orange) and terrace plateau region (TP region, blue). i) Lateral current profile (under +0.3 V tip bias) averaged along the dashed guidelines in (h). TEB. The error bars correspond to the standard deviations of the current distribution over the 24 pixels along the imaginary guidelines. j) Average *I-V* curves over the TP ($I_{TP}$-*V*, circle, blue) and TE region ($I_{TE}$-*V*, square, orange), and the corresponding fitting curves (solid lines) based on the Simmons tunneling model. The schematic barrier potential profile used to fit the $I_{TE}$-*V* and $I_{TP}$-*V* curves is shown in the right panel. For fitting the $I_{TP}$-*V* curve, the barrier height ($\varphi_B$) and width ($t_B$) were set as 0.375 eV and 1.2 nm (3 u.c.), respectively. For fitting the $I_{TE}$-*V* curve, $\varphi_B$ was reduced slightly to 0.370 eV, while $t_B$ was reduced to 0.98 nm.

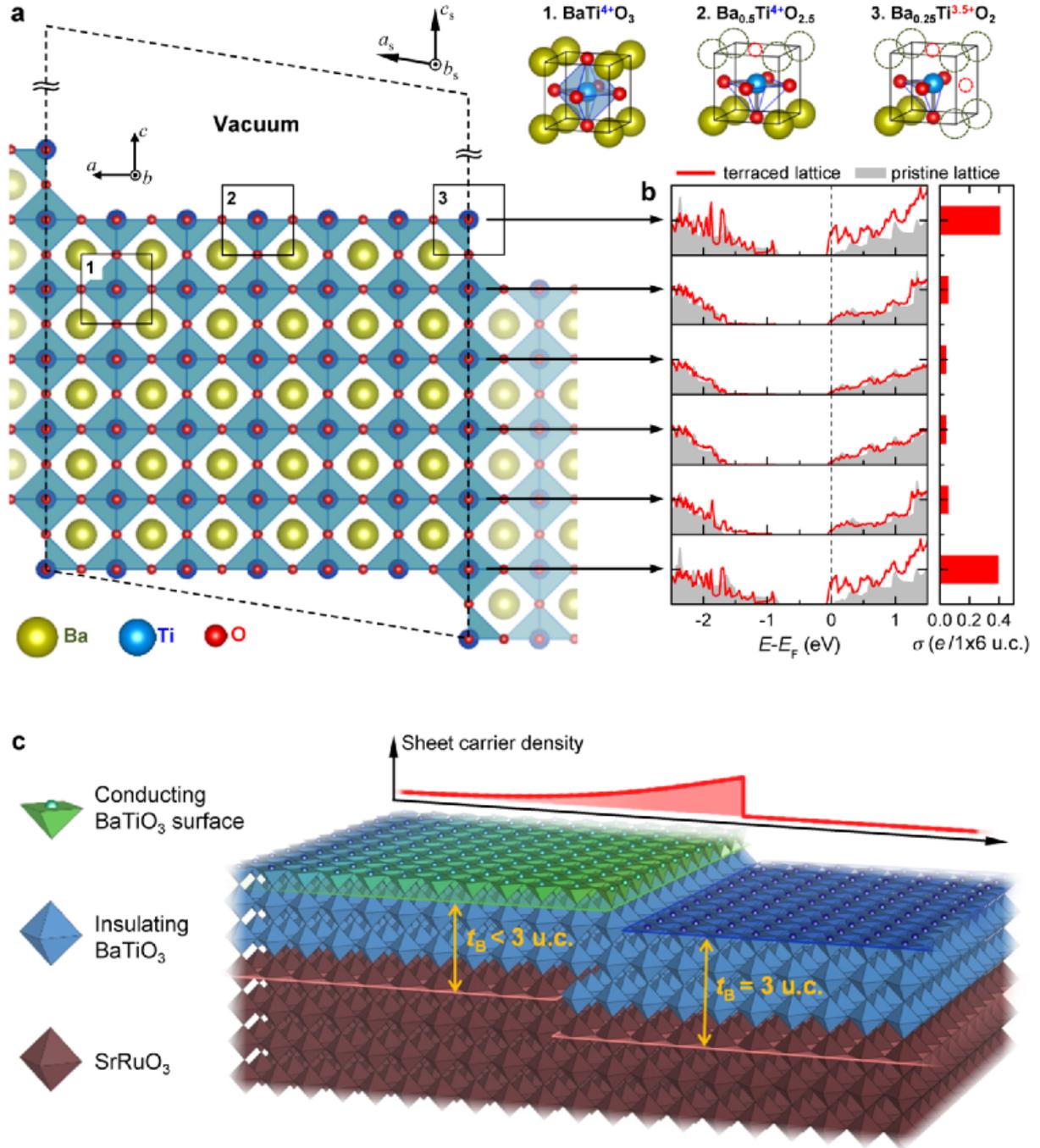

**Figure 3**. a) Schematic lattice of a BTO supercell constructed for first-principles calculations. The terraced structure is induced by setting the supercell vector $a_s = 6a + c$. In the supercell, both the surfaces and terrace edges are set to be $TiO_2$-terminated. The detailed atomic positions and ionic valence states in the BTO u.c. at different locations are also shown. The electronic reconstruction can occur at the terrace edge unit-cell and reduce the Ti valence to +3.5. b) The density of states (DOS) profiles and free electron densities of the $TiO_2$ planes in BTO supercells with (red) and without (gray) terrace structure. For $TiO_2$ planes at the terraced surface, the DOS of the Ti $3d$ band is enhanced near the Fermi level ($E_F$). The free electron densities $\sigma$ can reach ~0.4 $e$ per $W \times 1$ surface u.c.. c) Schematic illustration of the tunneling conductance enhancement at a terrace edge. As illustrated by the schematic sheet carrier density profile of the $TiO_2$ surface layer (thick solid lines, red), the free electrons spread over the TE region. Integration of these free electrons over $W \times 1$ surface u.c. should be ~ 0.5 $e$.

These free carriers make the surface layer at the TE region (green) become metallic, leading to a reduction in the effective $t_B$.

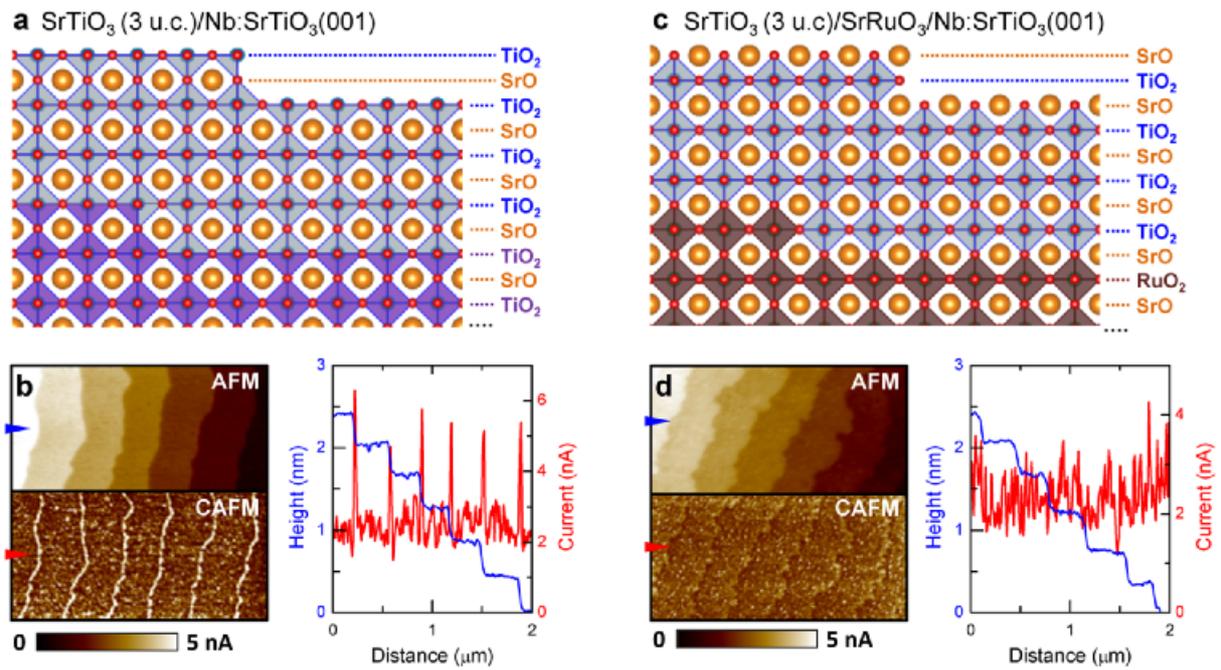

**Figure 4**. a,b) Schematic lattice of (a) STO(3 u.c.)/NSTO(001) and (b) STO(3 u.c.)/SRO/NSTO(001) films near a terrace edge. c,d) AFM topographic images (top panel) and CAFM images (bottom panel, under 0.5 *V* tip bias) of (c) STO(3 u.c.)/NSTO(001) and (d) STO(3 u.c.)/SRO/NSTO(001) films. The scanning area is $1 \times 2$ μm$^2$. AFM (CAFM) horizontal line profiles along the blue (red) triangles are shown in the right panel. The TiO$_2$-terminated STO/NSTO(001) film clearly exhibits higher tunneling conductance at terrace edges, while the SrO-terminated STO/SRO/NSTO(001) film only shows a weak current contrast probably due to topographic crosstalk.

# Supporting Information

**Electronic Reconstruction Enhanced Tunneling Conductance at Terrace Edges of Ultrathin Oxide Films**

*Lingfei Wang*[1,2], *Rokyeon Kim*[1,2], *Yoonkoo Kim*[3], *Choong H. Kim*[1,2], *Sangwoon Hwang*[1,2], *Myung Rae Cho*[1,2], *Yeong Jae Shin*[1,2], *Saikat Das*[1,2], *Jeong Rae Kim*[1,2], *Sergei V. Kalinin*[4], *Miyoung Kim*[3], *Sang Mo Yang*[4,5], *and Tae Won Noh*[1,2]

[1]Center for Correlated Electron Systems, Institute for Basic Science (IBS), Seoul 08826, Republic of Korea;
[2]Department of Physics and Astronomy, Seoul National University, Seoul 08826, Republic of Korea
[3]Department of Materials Science and Engineering and Research Institute of Advanced Materials, Seoul National University, Seoul 08826, Republic of Korea
[4]Center for Nanophase Materials Sciences, Oak Ridge National Laboratory, Oak Ridge, Tennessee 37831, United States
[5]Department of Physics, Sookmyung Women's University, Seoul 04310, Republic of Korea

## 1. Oxide film growth and structural characterizations

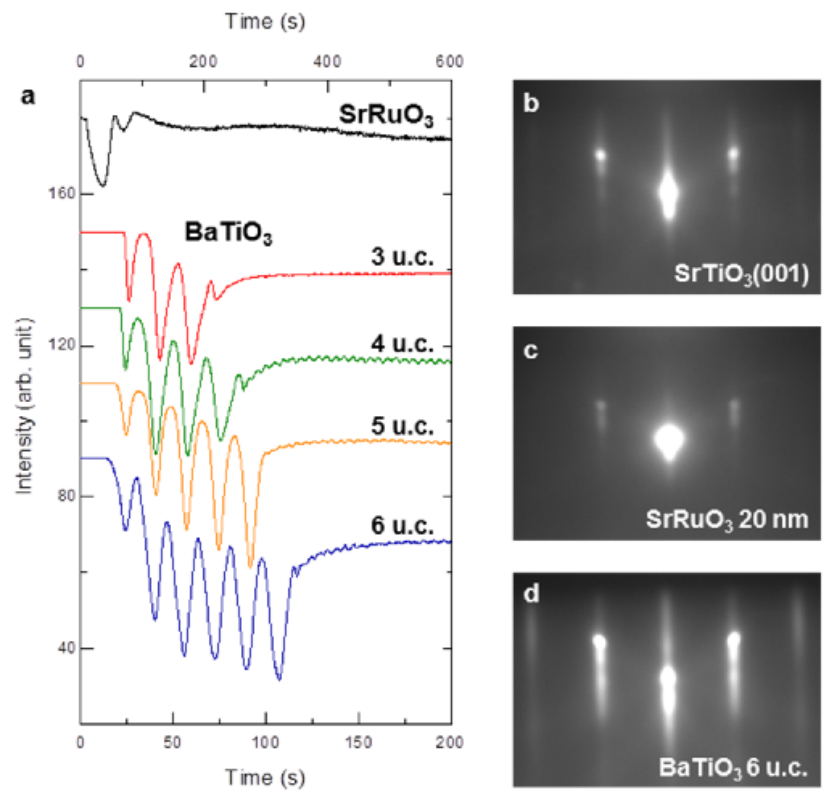

**Figure S1.** Epitaxial growth of the BTO/SRO/STO(001) films. a) Time-dependent RHEED intensity profiles of the specular spot during the SRO and BTO growth. b-d) RHEED diffraction patterns of (b) STO(001) substrate, (c) SRO(20 nm)/STO(001) film, and (d) BTO(6 u.c.)/SRO/STO(001) film. The RHEED profile of SRO growth exhibits a clear growth mode transition from layer-by-layer to step-flow.[1] Meanwhile, the surface termination was converted from $RuO_2$ to SrO.[2]

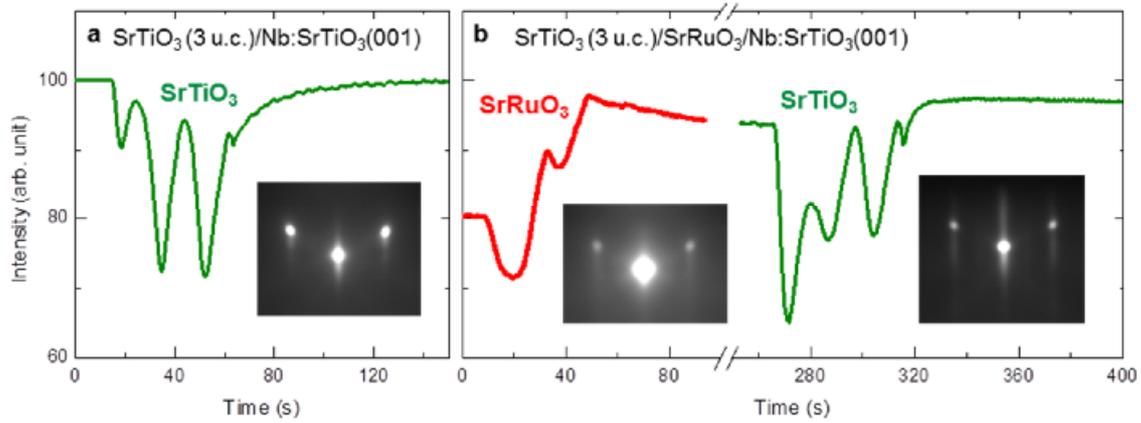

**Figure S2.** Epitaxial growth of the ultrathin STO films. a) Time-dependent RHEED intensity profile of the specular spot during the growth of 3 u.c. STO on NSTO(001) substrate. b) Time-dependent RHEED intensity profiles of the specular spot during the growths of 10 nm SRO and 3 u.c. STO on NSTO(001) substrate.

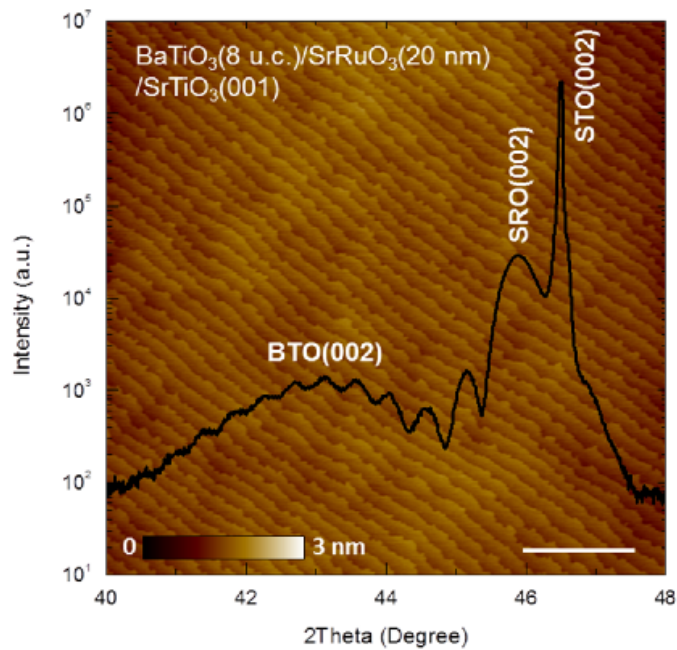

**Figure S3.** Structural Characterizations of BTO/SRO/STO(001) films. The X-ray diffraction $\omega$-$2\theta$ scan measured from the BTO(8 u.c.)/SRO/BTO film. The diffraction peaks were indexed in pseudo-cubic perovskite notation. The inset shows the AFM image with a scanning area of 15 × 15 μm$^2$. The scale bar corresponds to 3 μm. The clear Laue fringes in the $\omega$-$2\theta$ scan and uniform terrace structure in AFM image confirm the high epitaxial quality of the BTO film.

## 2. Scanning probe microscopy-based characterization techniques

**CAFM and spatially-resolved *I-V* spectroscopy.** Schematic illustrations of the CAFM and spatially-resolved *I-V* spectroscopy measurements are shown in Figure S4a. AFM tips (Bruker, DDESP-V2) with conductive-diamond coating and high spring constant (~80 nN/nm) were used for these measurements.[3] The robust conductive coating ensures the stability of current measurements. We fixed the contact forces between tip and sample at ~1000 nN (~500 nN) for CAFM (*I-V* spectroscopy) measurement via controlling the cantilever deflection, in order to minimize the current variation originated from poor tip/sample contact. During the *I-V* spectroscopy measurements, we superposed an $n \times m$ grid on a pre-scanned area of interest. The minimum pixel size, limited by the tip radius, was $10 \times 10$ nm$^2$. At each pixel, we applied a constant DC bias or a triangular DC voltage waveform to the tip (Figure S4a$_1$), and detected the current through SRO bottom electrode (Figure S4a$_2$) using a commercial lock-in amplifier (Femto, DLPCA-200).[4] Data recording and proceeding were done by custom-written LabVIEW and MATLAB programs.

**Band excitation-piezoresponse switching spectroscopy (BEPS).** Schematic illustrations of the BEPS measurements are shown in Figure S4b. Ir/Pt-coated AFM tips (Nanosensors, PPP-EFM) were used for this measurement. The band excitation (BE) waveform contains a band of frequencies near the cantilever resonance (inset of Figure S4b$_2$). During the BEPS measurements, we superposed the BE waveform on a bipolar triangular switching waveform with a series of DC poling voltage ($V_{dc}$) pulses (Figure S4b$_2$). We applied this waveform to the tip for perturbing the sample. Then the response of the cantilever can be measured and Fourier-transformed simultaneously to give the piezoresponse signal. Using a simple harmonic oscillator model fitting, we can extract the piezoresponse signals [resonance amplitude ($A_0$), phase ($\theta$), resonance frequency ($\omega_0$) and quality factor ($Q$)] during (ON field signal) or after (OFF field signal) each $V_{dc}$ pulses (Figure S4b$_1$). Using such band-excitation technique, we can make sure all the piezoresponse signals were collected at the resonance frequency even when it changes with tip-surface interaction. Therefore, we can minimize the crosstalk from topography on piezoresponse signal.[5]

We also performed the BEPS measurement pixel by pixel in a preselected area of interest, to construct a multidimensional mapping of ferroelectricity in BTO ultrathin films.[6]

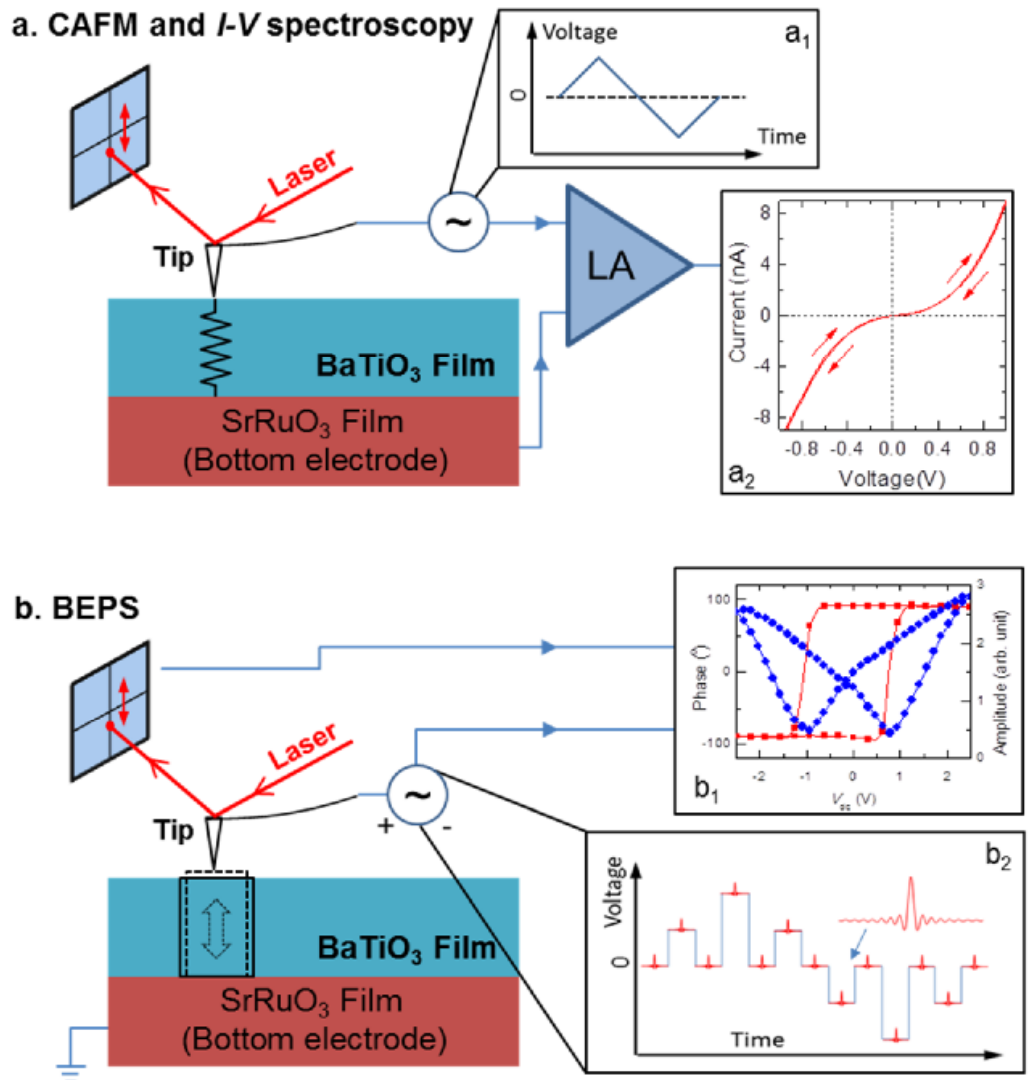

**Figure S4.** a,b) Schematics of various scanning probe microscopies, including (a) CAFM and spatially-resolved *I-V* spectroscopy, (b) BEPS.

## 3. Characterizations of ferroelectricity

We characterized the ferroelectricity of the BTO ultrathin films with $t_{BTO}$ = 3, 5 and 8 u.c. by BEPS technique (See Section 2). Figure S5 shows the OFF-field piezoresponse amplitude ($A_0$) and phase ($\theta$) signals as a function of the $V_{dc}$ pulse. Both the $\theta-V_{dc}$ hysteresis loops and $A_0-V_{dc}$ butterfly curves clearly exhibit FE-like feature. We also performed the contact Kelvin probe force microscopy measurements (not shown here) to exclude the non-FE mechanisms of the observed BEPS results.[7-10] Based on these results, we conclude that all the BTO ultrathin films are FE. Note that the piezoresponse amplitude decreases significantly as the $t_{BTO}$ decreases to 3 u.c.. This feature indicates that the FE polarization reduces substantially when the $t_{BTO}$ is approaching the FE critical thickness.

We also carried out the spatially-resolved BEPS mapping over a 50 × 50 grid on an area of 500 × 500 nm$^2$ on the 3 u.c. BTO sample.[6] As shown in Figure S6, the amplitude and phase signals of the piezoresponse are homogeneous over the preselected region. The resonance frequency shows a decrease of ~2 kHz at the terrace edges, which could originate from the tip-surface spring constant change induced by the sharp topographic change at the terrace edges.[11]

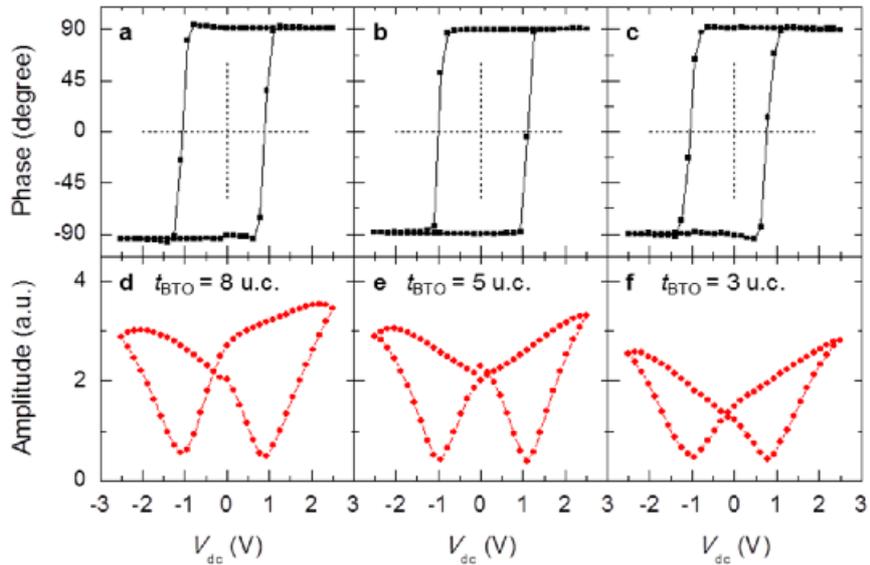

**Figure S5.** Ferroelectric characterizations in BTO/SRO/STO(001) films with various $t_{BTO}$. a-f) $V_{dc}$ dependent OFF-field (a-c) phase and (d-f) amplitude of the piezoresponse, which are measured from BTO films with various $t_{BTO}$.

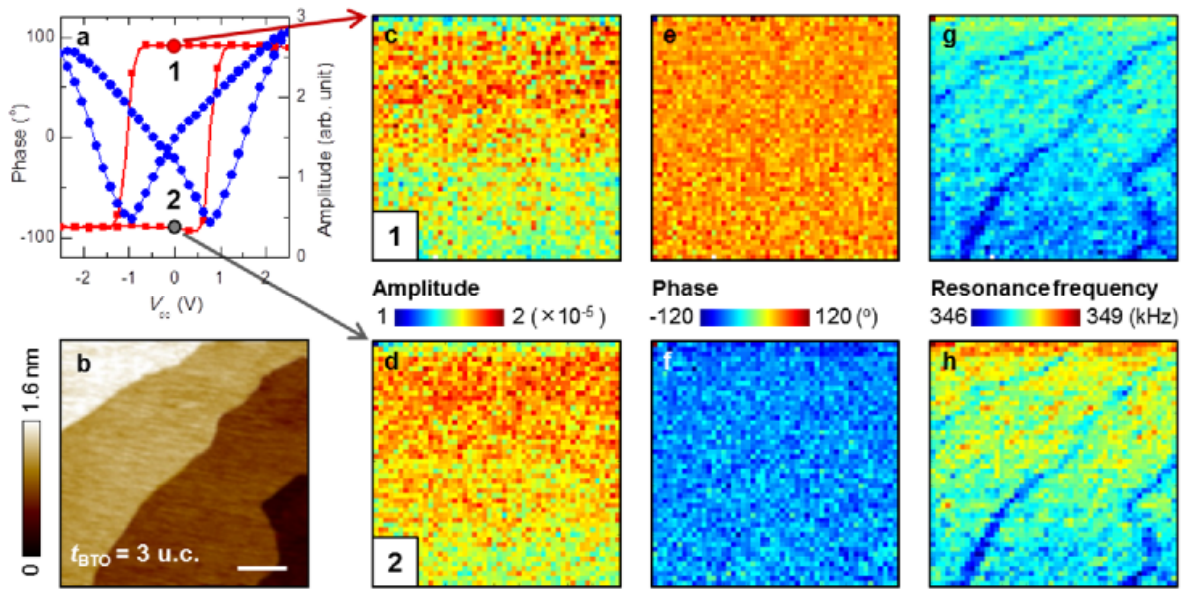

**Figure S6.** Spatially-resolved BEPS of 3 u.c. BTO film. a) $V_{dc}$ dependent phase and amplitude of the piezoresponse. b) AFM topographic image of a preselected area (500 × 500 nm$^2$). The scale bar refers to 100 nm. c-h) Spatially-resolved mappings of the (c)[(d)] amplitudes, (e)[(f)] phases, and (g)[(h)] resonance frequencies recorded at $V_{dc}$ = 0 with the upward (downward) polarization. These two polarization states were marked by point 1 and 2 in (a). Note that the gradual changes of amplitude from top to bottom in (c) and (d) originated from a slight tip/surface contact condition variation during measurements.

## 4. *I-V* characterizations of the ultrathin BTO-based FTJ

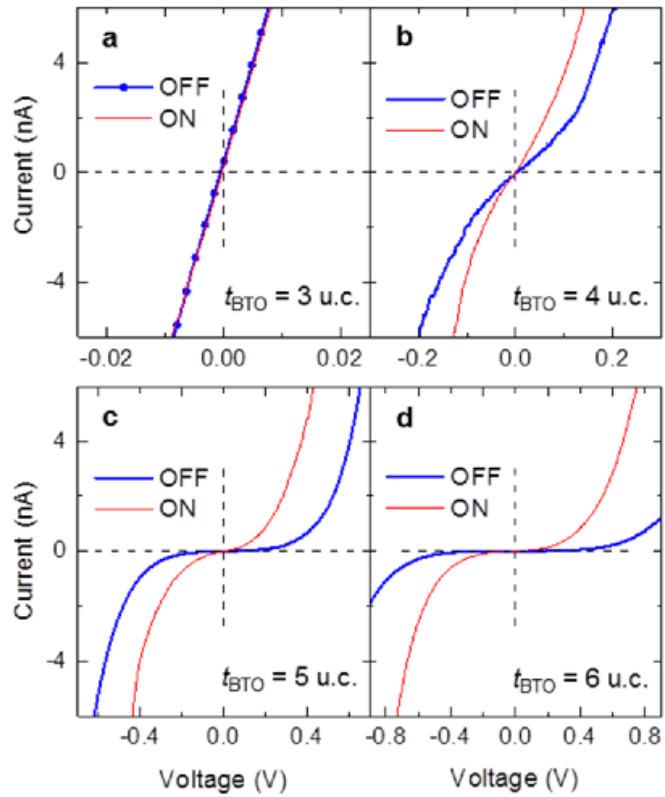

**Figure S7.** a-d) *I-V* curves of the ultrathin BTO-based FTJ with various $t_{BTO}$. For the FTJ with $t_{BTO}$ = 3 u.c., the $I_{ON}$-*V* and $I_{OFF}$-*V* curves are linear and almost identical, indicating a negligible TER effect. As $t_{BTO}$ increases above 4 u.c., the $I_{ON}$-*V* and $I_{OFF}$-*V* curves become more nonlinear and different, signifying a gradually enhanced TER effect.

## 5. Tunneling mechanisms and fitting results of the *I-V* curves

Fitting the experimental *I-V* curves based on proper tunneling models can help us to obtain more information about the tunneling barrier. The current flows through ferroelectric or dielectric thin films can be mainly ascribed into three transport mechanisms: direct tunneling (DT), Fowler-Nordheim tunneling (FNT), and thermionic emission (TI). [12]

First, we exclude the TI mechanism based on the following reasons. According to the theoretical paper from Pantel et al.,[13] TI only dominates the conduction when the FE film is thicker than 3 nm, and the *I-V* curve should be highly asymmetric. These features are quite different from those observed in our ultrathin BTO samples (only ~1.2 nm thick, and *I-V* curves are nearly symmetric). Moreover, in the previous experimental reports on the FTJ, the TI mechanism only becomes applicable for two kinds of device geometry: 1) FTJ with barriers up to 9 nm thick,[14] and 2) FTJ containing semiconductor electrodes, which can form additional barrier due to carrier depletion.[15,16] All the other FTJ with conventional metal/ferroelectric/metal structures and thinner barriers, to the best of our knowledge, were described by DT and/or FNT.[12] Therefore, in the following analyses and discussions, we will consider the DT and FNT mechanisms only.

For typical FTJ under zero bias, the electrostatic potential profiles of FE barrier above Fermi level usually have a trapezoidal shape.[3,17] When the FE polarization becomes weak, we can simplify the trapezoidal barrier by a rectangular barrier with the average potential height.[18] We assume that the average barrier height is $\varphi_B$ and the barrier width is $t_B$. When the external bias is small ($|V| \leq 2\varphi_B/e$), the conduction through the BTO barrier is dictated by DT mechanism. In this case, the potential profile under external bias *V* can be expressed as

$$\varphi(x, V) = \varphi_B + \left(\frac{1}{2} - \frac{x}{t_B}\right)eV$$

Using the WKB approximation, the current density (*J*) from DT is given by the Simmons equation,[19]

$$J(V) = \frac{e}{4\pi^2\hbar t_B^2}\left\{\left(\varphi_B - \frac{eV}{2}\right)\exp\left[-\frac{2(2m_e^*)^{1/2}}{\hbar}\left(\varphi_B - \frac{eV}{2}\right)^{1/2}t_B\right]\right.$$
$$\left. - \left(\varphi_B + \frac{eV}{2}\right)\exp\left[-\frac{2(2m_e^*)^{1/2}}{\hbar}\left(\varphi_B + \frac{eV}{2}\right)^{1/2}t_B\right]\right\}$$

(1)

where $\hbar$ is the reduced Planck constant, $m_e^*$ is the effective electron mass. Note that this equation only valid when $|V| \leq 2\varphi_B/e$.

Under a large bias ($|V| > 2\varphi_B/e$), the barrier shape becomes triangular. In this case, the current density is dictated by the FNT,[13]

$$J(V) = \frac{e^3 m_e}{16\pi^2 \hbar m_e^* \varphi_B} \left(\frac{V}{t_B}\right)^2 \exp\left[-\frac{4(2m_e^*)^{1/2}}{3\hbar e}\frac{\varphi_B^{3/2} t_B}{V}\right]$$

(2)

where $m_e$ is the free electron mass.

We employed equation (1) to fit the $I_{ON}$-$t_{BTO}$ and $I_{OFF}$-$t_{BTO}$ curves at 0.4 V (fixing the $\varphi_B$ values for ON and OFF states and assuming the $t_B = t_{BTO}$).[3] Then we can calculate and plot the $t_{BTO}$ dependent ON/OFF ratio curve (dotted line in Figure 1f). This curve can be used as a reference for addressing the FTJ performance degradation.

We also employed the equation (1) to fit the $I_{TP}$-$V$ and $I_{TE}$-$V$ curves (Figure 2j) under small bias ($|V| \leq 2\varphi_B/e \approx 0.7$ V). There are three variable fitting parameters in equation (1): barrier height $\varphi_B$, barrier width $t_B$ and effective electron mass $m_e^*$. We first fit the $I_{TP}$-$V$ curve by assuming $t_B = t_{BTO} = 1.2$ nm (3 u.c. BTO) and varying $\varphi_B$ and $m_e^*$. Good fitting can be obtained with $\varphi_B = 0.375$ eV (Figure S8a), which is consistent with the average barrier height reported previously.[3] And the optimum $m_e^* = 1.05\ m_e$ is quite close to the free electron mass.

We tried two ways to fit the $I_{TE}$-$V$ curves under small bias, which are described as follows. All the fitting equations and parameters were summarized in Table S1.

**Method 1.** We initially assumed that the larger current flow through the TE region can be interpreted by an additional leakage channel with Ohmic $I$-$V$ behavior under small bias,[20] connected in parallel with the tunneling conduction channel. So we fitted the $I_{TE}$-$V$ curve by $I_{TE}(V) = I_{TP}(V) + V/R_L$, where the $R_L$ is the effective resistance of leakage conduction. The optimized $R_L$ value is $3.25 \times 10^8$ Ω. The fitting result is shown in Figure S8c.

**Method 2.** We fit the $I_{TE}$-$V$ curve by varying the tunneling barrier parameters ($t_B$ and $\varphi_B$) and $m_e^*$ used for fitting the $I_{TP}$-$V$ curve. To achieve the optimum fitting results, the $t_B$ have to reduce by approximately 0.5 u.c. (from 1.20 to 0.98 nm), while $\varphi_B$ ($m_e^*$) only changes slightly from 0.375 eV (1.05 $m_e$) to 3.70 eV (1.02 $m_e$). The fitting result is shown in Figure 3j and Figure S8d.

In order to compare these two fitting methods, we calculated the mean relative errors (MRE) by the equation

$$\mathrm{MRE} = \frac{1}{N}\sum \left|[I_{\mathrm{exp}}(V) - I_{\mathrm{fit}}(V)]/I_{\mathrm{fit}}(V)\right|$$

where $I_{\mathrm{exp}}(V)$ and $I_{\mathrm{fit}}(V)$ are the current values from experiments and fitting curves, respectively. As summarized in Table S1, the MRE calculated from Method 2 (2.68%) is much smaller than that from Method 1 (12.98%). For Method 1, the deviations between $I_{\mathrm{exp}}(V)$ and $I_{\mathrm{fit}}(V)$ can be clearly observed in small bias range (Figure S8c). Based on this analysis, we assert that the Method 2 can manifest the real conduction mechanism at the TE region. The higher conductance is mainly originated by the local tunneling barrier width reduction rather than additional leakage conduction channels.

In the large bias range ($|V| > 0.7$ V), we employed the equation (2) to fit the $I_{\mathrm{TP}}$-$V$ and $I_{\mathrm{TE}}$-$V$ curves based on FNT model. As shown in Figure S8b and S8e, using the same parameters obtained from the DT fitting (Table S1), we can reproduce the experimental curves well. These fitting results clearly demonstrate that the dominating conduction mechanisms in both TE and TP regions are DT in the small bias range and FNT in the large bias range.

| | | Fitting Equation | Fitting Parameters | | | | MRE [%] |
|---|---|---|---|---|---|---|---|
| | | | $t_{\mathrm{B}}$ [nm] | $\varphi_{\mathrm{B}}$ [eV] | $m_{\mathrm{e}}^*/m_{\mathrm{e}}$ | $R_{\mathrm{L}}$ [Ω] | |
| $I_{\mathrm{TP}}$-$V$ | | Simmons Equation | 1.20 | 0.375 | 1.05 | - | 3.26 |
| $I_{\mathrm{TE}}$-$V$ | Method 1 | $I_{\mathrm{TE}}(V) = I_{\mathrm{TP}}(V) + V/R_{\mathrm{L}}$ | 1.20 | 0.375 | 1.05 | 3.25×10⁸ | 12.98 |
| | Method 2 | Simmons Equation | 0.98 | 0.370 | 1.02 | - | 2.68 |

**Table S1.** Equations and parameters used for fitting the $I_{\mathrm{TP}}$-$V$ and $I_{\mathrm{TE}}$-$V$ curves.

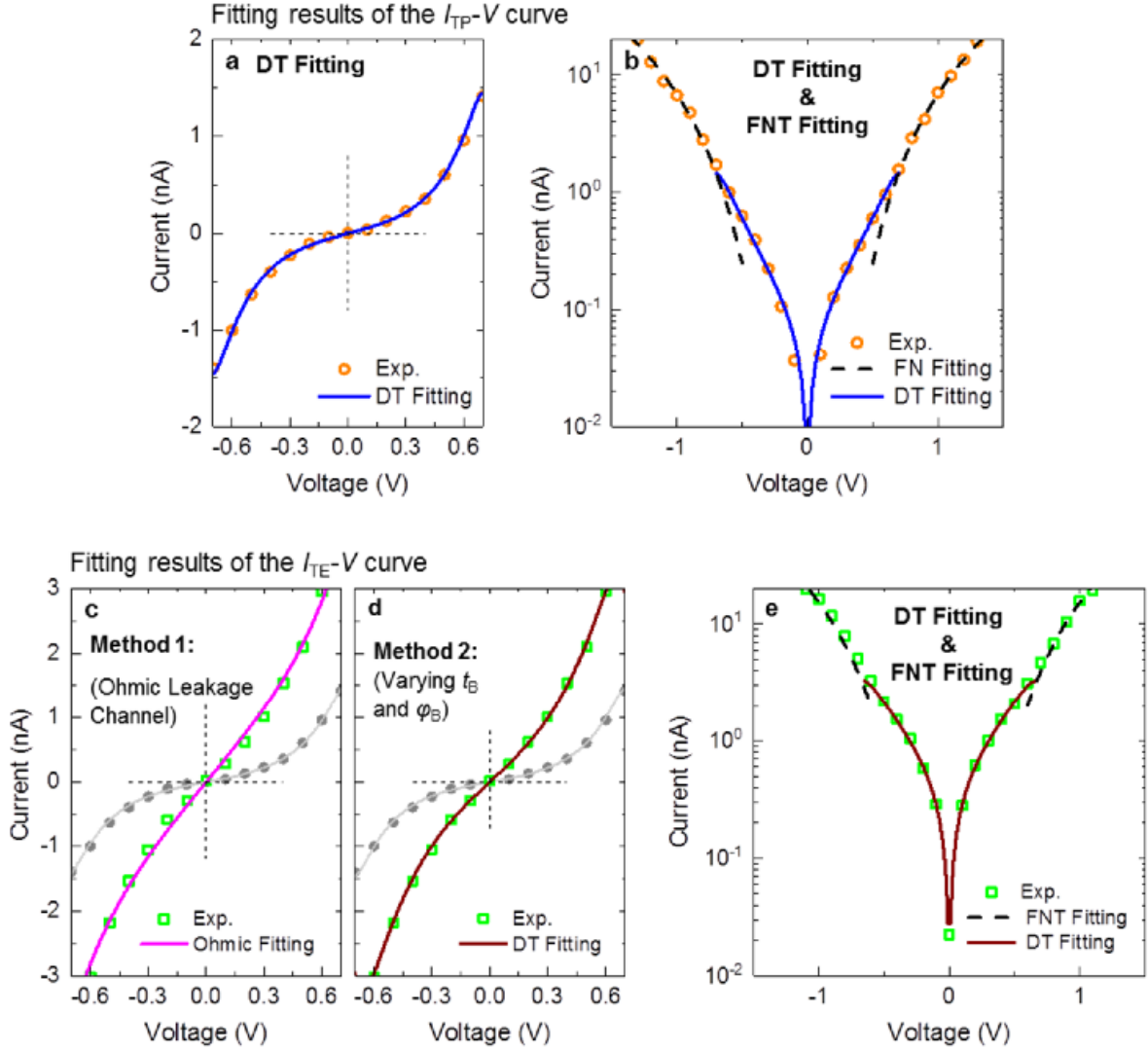

**Figure S8.** DT and FNT fitting results of the $I_{TP}$-$V$ and $I_{TE}$-$V$ curves measured from the 3 u.c. BTO sample. a) Linear plot of the experimental $I_{TP}$-$V$ curve and fitting curve in the small bias range ($|V| \leq 0.7$ V). In this bias range, the fitting is based on DT model (Simmons equation). b) Logarithmic plot of the experimental $I_{TP}$-$V$ curve and fitting curves. In the large bias range ($|V| > 0.7$ V), the fitting is based on FNT model. c,d) Linear plot of the experimental $I_{TE}$-$V$ curve and fitting curves in the small bias range (($|V| \leq 0.7$ V). The fitting in (c) is based on the equation $I_{TE}(V) = I_{TP}(V) + V/R_L$, and the fitting in (d) is based on the DT model. The $I_{TP}$-$V$ fitting results are inserted for comparison. e) Logarithmic plot of the experimental $I_{TE}$-$V$ curve and fitting curves. In the large bias range ($|V| > 0.7$ V), the fitting is based on FNT model.

## 6. Results and discussions for excluding the extrinsic effects.

Before exploring the origin of the local tunneling barrier variation, we would like to exclude several extrinsic effects that may also result in larger current across the terrace edges.

**1) Topographic crosstalk.**

During CAFM measurements, the tip may touch both the flat film surface and the atomic step at terrace edge. The larger contact area can increase the detected current. However, the tip radius (~10 nm) is much larger than the terrace height (~0.4 nm). Consequently, the contact area change should be limited to several percentages, which is far less than the observed $I_{peak}/I_{background}$ ratio (more than 3.0 for 3 u.c. BTO sample at $V_{read}$ = 0.5 V). Moreover, the current enhancement induced by contact area change should be a constant factor, which cannot explain the observed $t_{BTO}$-dependent (Figure 1) as well as the tip bias-dependent (Figure S9) conductance enhancement at the terrace edge ($I_{TE}/I_{TP}$ ratio). In addition, as shown in Figure 4, we can effectively control terrace edge conductance in STO ultrathin films by surface termination engineering. This result further demonstrated that the current contrast from topographic crosstalk is negligible compared with that from intrinsic electronic reconstruction.

**2) The lateral misalignment between the top and bottom terrace edges.**

As depicted in Figure S10, along the miscut direction, the top terrace edge could locate in front of or behind the bottom one, denoted as terrace geometry A or B, respectively. For the terrace geometry A (B), terrace misalignment makes the film one unit-cell thicker (thinner) and thus leads a smaller (larger) tunneling current at the terrace edge. According to the cross-sectional high-angle annular dark field (HAADF) image shown below (Figure S13), we can observe the terrace geometry A only in the BTO films. Note that the terrace misalignment according to TEM images is approximately 8 ± 3 nm, which is much thinner than the width of the highly conductive TE region. Accordingly, the terrace misalignment should neither contradict to nor trigger the local tunneling barrier variation near terrace edges.

**3) Oxygen vacancies and surface adsorbates accumulated at the terrace edges.**

According to Li *et al.*,[21] the oxygen vacancies may preferentially accumulate at the terrace edges and increase the local surface conduction. In this case, the conductance enhancement near the terrace edges should highly depend on the ex-situ annealing conditions. In order to exclude the possible conduction from oxygen vacancies, we ex-situ annealed our BTO films at 600 ºC in ambient oxygen flow for 1 and 5 h, and then performed the CAFM measurements. As shown in Figure S11, the enhanced conductance at terrace edges does not show any change for these two ex-situ annealing conditions. These results indicate that the BTO films were well oxidized already after 1h annealing.

The polar adsorbates in the ambient environment (e.g., $H^+$ or $OH^-$) may also change the surface conductance by varying the electrostatic boundary conditions.[22] To exclude this effect, we performed the CAFM measurements in an AFM system (Cypher, MFP-3D, Asylum Research) installed in a vacuum glove box. The samples were annealed in vacuum (0.1 mTorr) at 250 and 350 ºC for 1 h to remove the surface adsorbates.[23] Then the samples were cooled down to room temperate for in-situ AFM and CAFM measurements. During the all the steps the sample was kept in the vacuum environment. As shown in Figure S12, after annealed in vacuum at 250 and 350 ºC, the CAFM images (Figure S12b and S12c) do not show any obvious change compared to the ambient counterpart (Figure S12a).

Therefore, we believe that the preferential distribution of oxygen vacancies and surface adsorbates should not be the origin of highly conductive terrace edges.

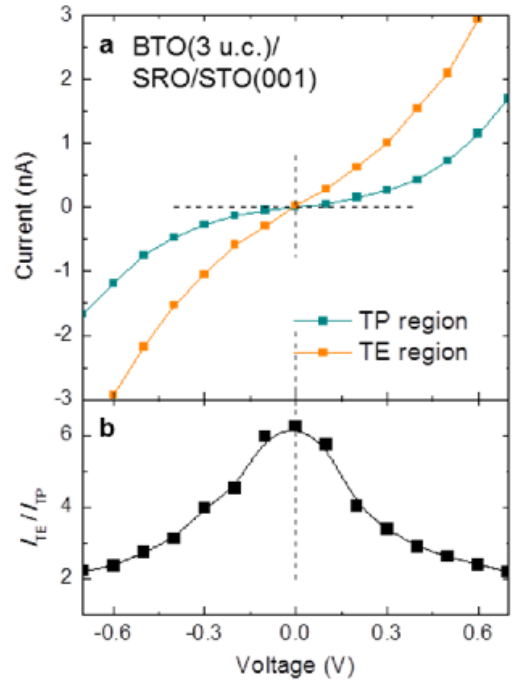

**Figure S9.** a) Experimental $I_{TE}$-$V$ and $I_{TP}$-$V$ curves. b) Tip bias-dependent $I_{TE}/I_{TP}$ ratio.

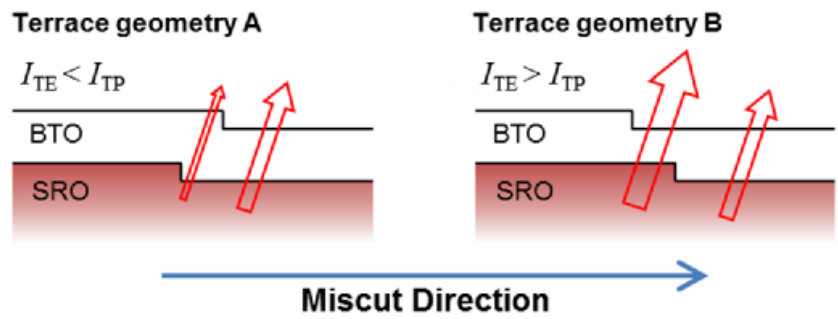

**Figure S10.** Schematic illustrations of the two possible geometries of terrace edge mismatch. The current flows through BTO barriers are schematically illustrated by the open (red) arrows. The miscut direction is marked by the solid (blue) arrow.

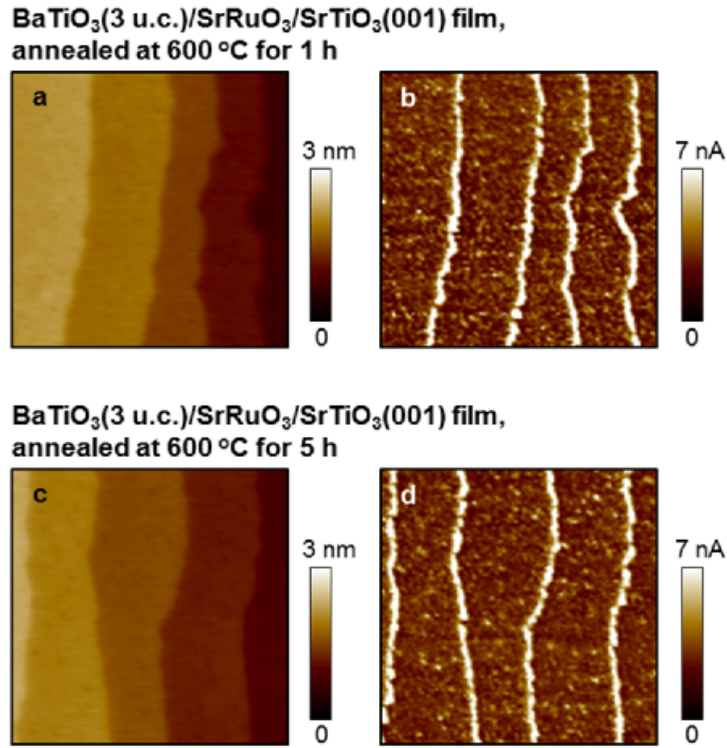

**Figure S11.** a-d) AFM and CAFM images measured from 3 u.c. BTO samples with various ex-situ oxygen annealing processes. (a) [(c)] AFM and (b) [(d)] CAFM images measured from the sample annealed at 600 °C for 1 h (5 h) in ambient oxygen flow. The scanning area is 1 × 1 µm$^2$. During the annealing processes, the sample does not show any obvious changes in both morphology and tunneling conductance.

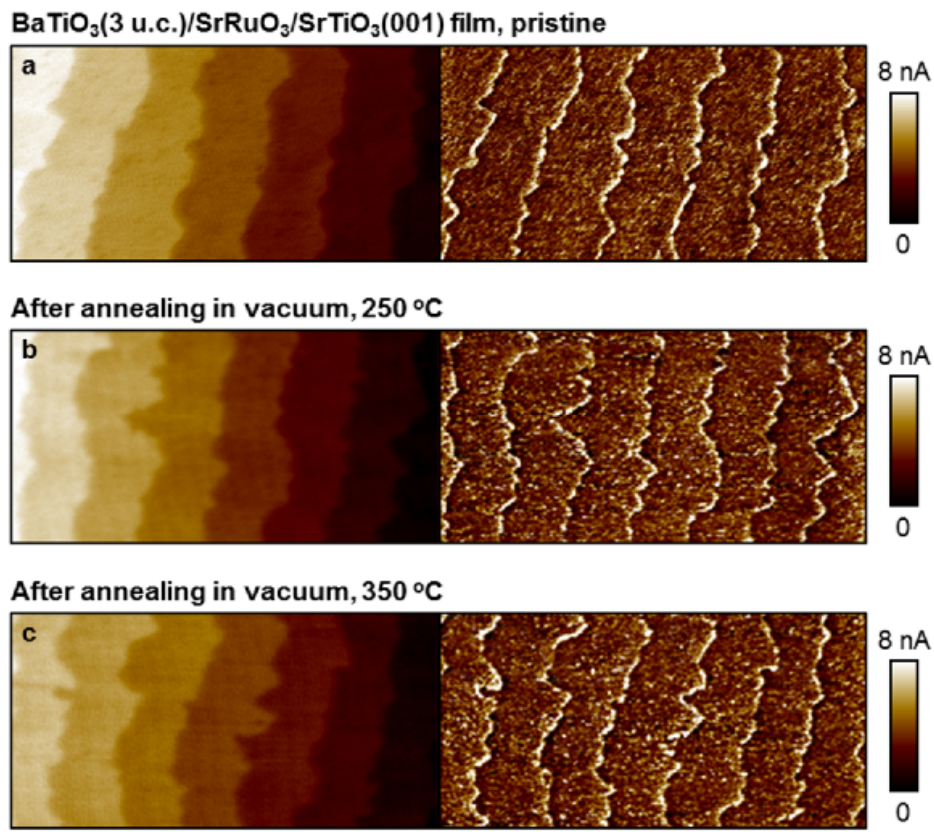

**Figure S12** a-c) AFM and CAFM images measured from 3 u.c. BTO samples under 3 conditions: (a) before annealing, after annealing in vacuum (~0.1 mTorr) at (b) 250 °C and (c) 350 °C for 1 h. The scanning area is 1 × 2 μm$^2$. During the annealing processes, the sample does not show any obvious changes in both morphology and tunneling conductance.

## 7. Scanning transmission electron microscopy

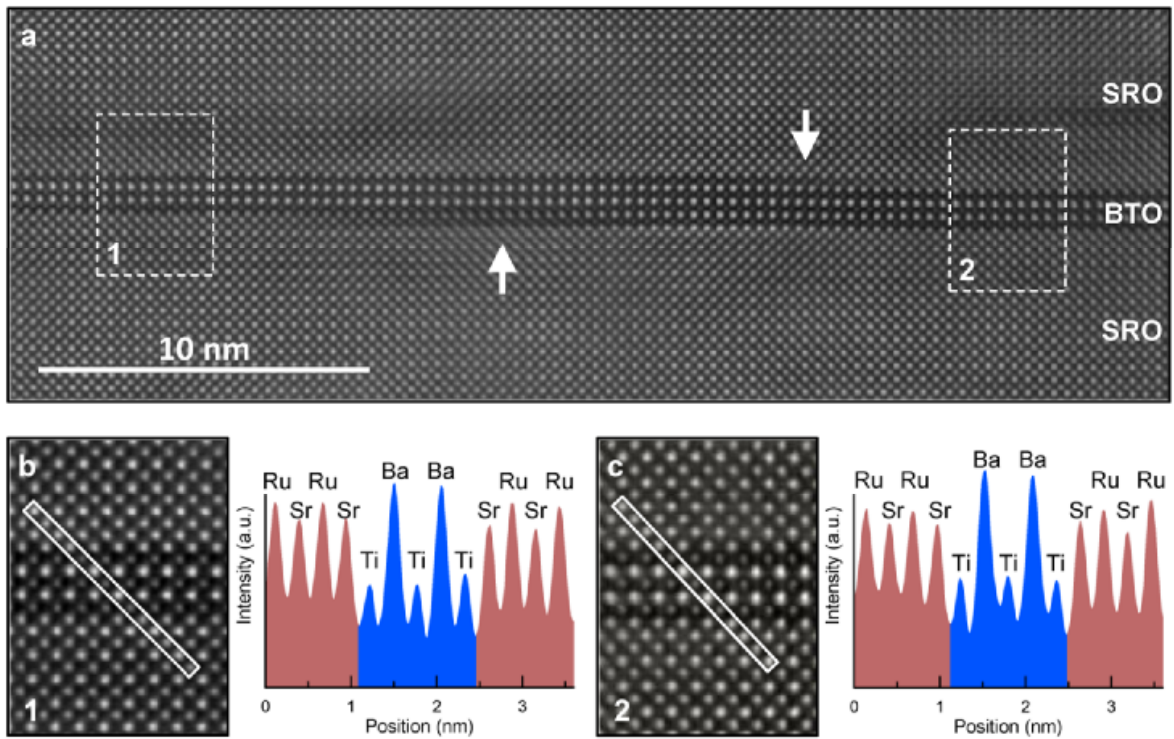

**Figure S13.** HAADF images of the SRO/BTO(3 u.c.)/SRO/STO(001) heterostructure near a terrace edge. a) Large scale HAADF images viewed along [100] axis over a terrace edge. The scale bar is 10 nm. The top and bottom terrace edges are marked by the solid arrows (white). Note that the terrace edges in this image are not laterally sharp in atomic scale because the terraces are not well aligned along the {100} crystallographic axes. b,c) Magnified HAADF images obtained in the dashed white boxes 1 and 2 in (a). The corresponding line profiles of HAADF intensity along the highlighted atomic columns [marked by solid (white) boxes in (b) and (c)] were also shown. The elements were identified and labeled in these profiles. In both regions, the bottom and top SRO/BTO interfaces are terminated by SrO/TiO$_2$ sequence.

Note that we use an SRO/BTO/SRO/STO(001) capacitor sample instead of the BTO/SRO/STO(001) film for STEM measurements. The SRO capping layer is necessary to prevent any damage on the ultrathin BTO during the STEM specimen preparation.

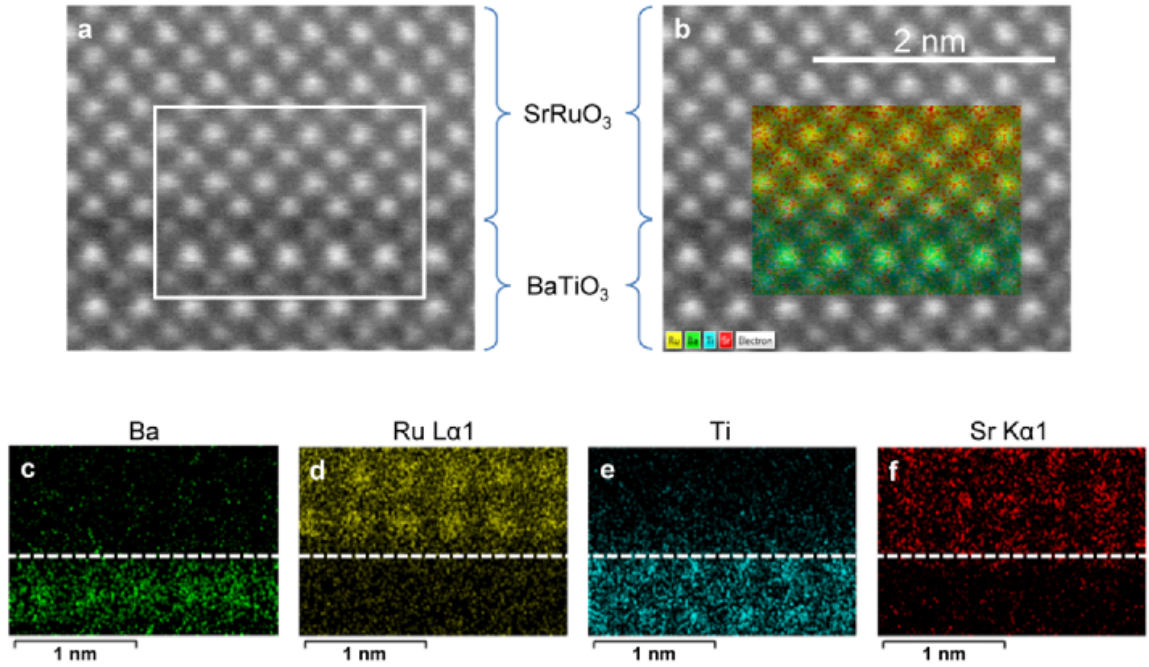

**Figure S14.** Energy dispersive spectroscopy (EDS) images near the top SRO/BTO interface of the SRO/BTO(3 u.c.)/SRO/STO(001) heterostructure. a) Magnified HAADF image (viewed along [100] axis) measured at the top SRO/BTO interface region. The preselected region for EDS imaging is marked by a solid (white) box. b) A overlapped EDS image of Ba, Ru, Ti and Sr elements. This image confirmed that the top SRO/BTO interface is atomically sharp without any inter-diffusion. c-f) Atomically-resolved EDS maps of (c) Ba, (d) Ru, (e) Ti and (f) Sr. The cut-off edges of Ba and Ru EDS intensities have a small distance away from the SRO/BTO interface (marked by the dashed line), while the Sr and Ti EDS maps exhibit sharp changes right at the heterointerfaces. These results further confirmed the SrO/TiO$_2$ interfacial termination sequence.

According to these HAADF images and EDS images, we demonstrated that both the bottom and top SRO/BTO interfaces are uniformly terminated by SrO/TiO$_2$ sequence. During the film growth, due to the high volatility of RuO$_2$ compounds, the SRO surface always has a SrO termination. The subsequently deposited BTO follows a unit-cell-by-unit-cell growth mode, and the surface is expected to be BaO-terminated. However, our recent results indicate that BaO is highly unstable under our growth condition (5 mTorr, 700 ºC), leading to a uniformly TiO$_2$-terminated surface.[24] On this basis, we believe atomic termination of the terrace edge, albeit being directly imaged, should be the same as the flat surface, i.e. TiO$_2$.

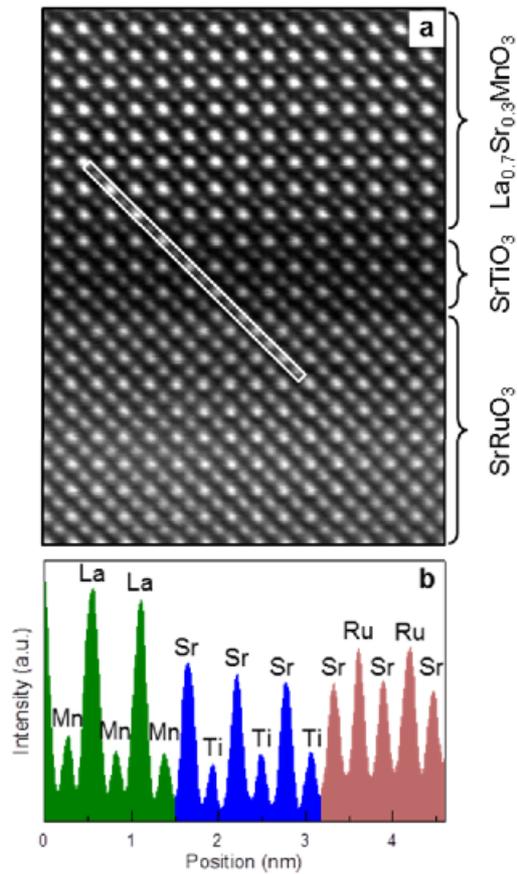

**Figure S15.** HAADF images of the $La_{0.7}Sr_{0.3}MnO_3$(20 u.c.)/STO(3 u.c.)/SRO/STO(001) heterostructure. a) HAADF images viewed along [100] axis. The $La_{0.7}Sr_{0.3}MnO_3$ capping layer was deposited to protect the STO film surface. b) The line profile of HAADF intensity along the highlighted solid (white) box. The elements were identified and labeled in these profiles. In the bottom and top interfaces, the STO film exhibits $SrO/TiO_2$ and $SrO/MnO_2$ termination sequences, respectively.

## 8. Extended first-principles calculations

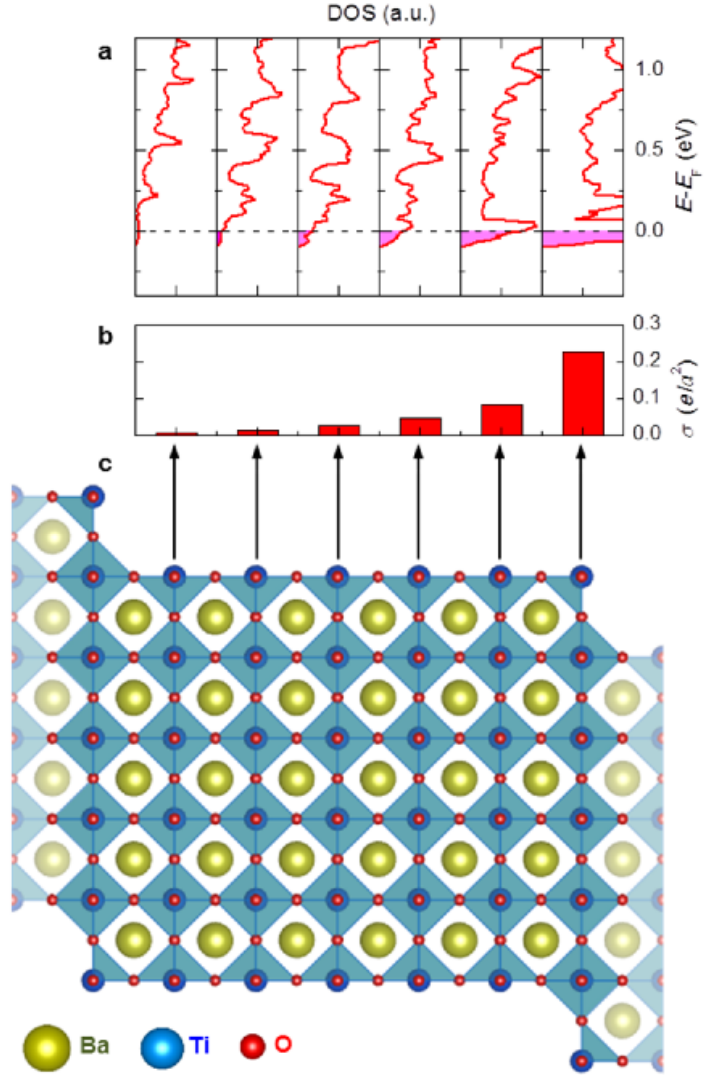

**Figure S16.** a,b) Local DOS profiles of Ti 3$d$ band [(a)] and carrier density [(b)] on each surface Ti site in the terraced BTO supercell. The occupied states below $E_F$ and carrier densities decrease gradually from the Ti cation at the terrace edge to the ones in the terrace plateau. c) Schematic lattice of BTO supercell used for DFT calculation.

**Estimate the carrier density at TE region.** According to Figure 2i, the width of conductive TE region ($W_{TE}$) is approximately 40 nm. We assumed that all the free electrons originate from electronic reconstruction are confined within the 0.5 u.c.-thick surface layer. The average carrier density $n_{ave}$ is estimated by

$$n_{ave} = \frac{0.5\,e}{0.5c \times a \times W_{TE}} = 1.6 \times 10^{20}\,(\text{cm}^{-3})$$

where *a* and *c* are the in-plane and out-of-plane lattice constants of BTO. Here we set $a = c = 0.4$ nm for simplification.

According to T. Kolodiazhnyi et al.,[25] the estimated carrier density of $\sim 10^{20}$ cm$^{-3}$ is high enough to make the 0.5 u.c. thick surface layer metallic. Moreover, with this free carrier density, BTO can still retain the low-symmetry structure. Therefore, we can observe substantially enhanced tunneling conductance but negligible change in the piezoresponse (Figure S6) near terrace edges.

## 9. Extended spatially-resolved *I-V* spectroscopy results

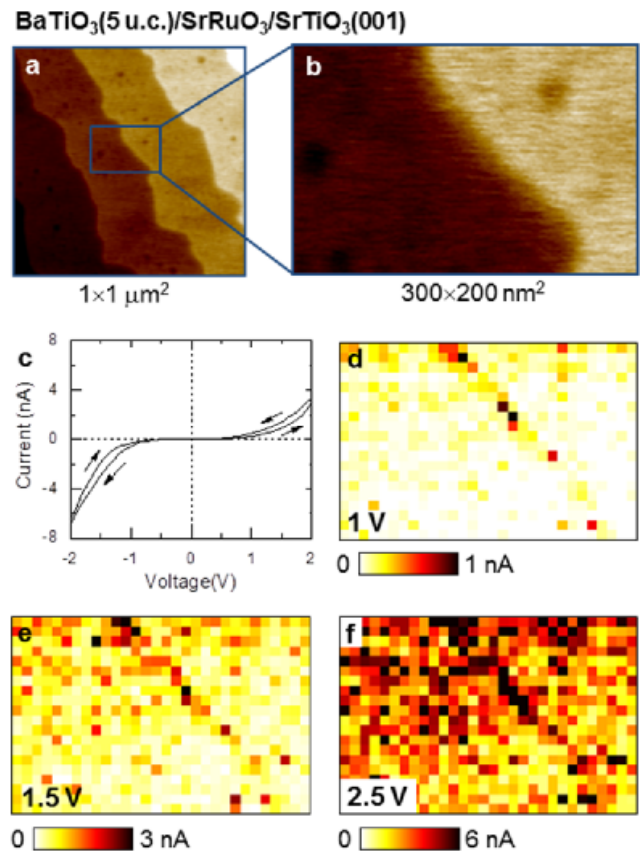

**Figure S17.** a,b) AFM topographic images of the 5 u.c. BTO film with a scanning area of (a) 1 × 1 μm² and (b) 300 × 200 nm². c) *I-V* curve averaged over the whole 30 × 20 pixels. d-f) Spatial maps of local current obtained at different tip biases. Comparing to the 3 u.c. BTO sample, all the maps exhibit much weaker conductance enhancement at terrace edges.

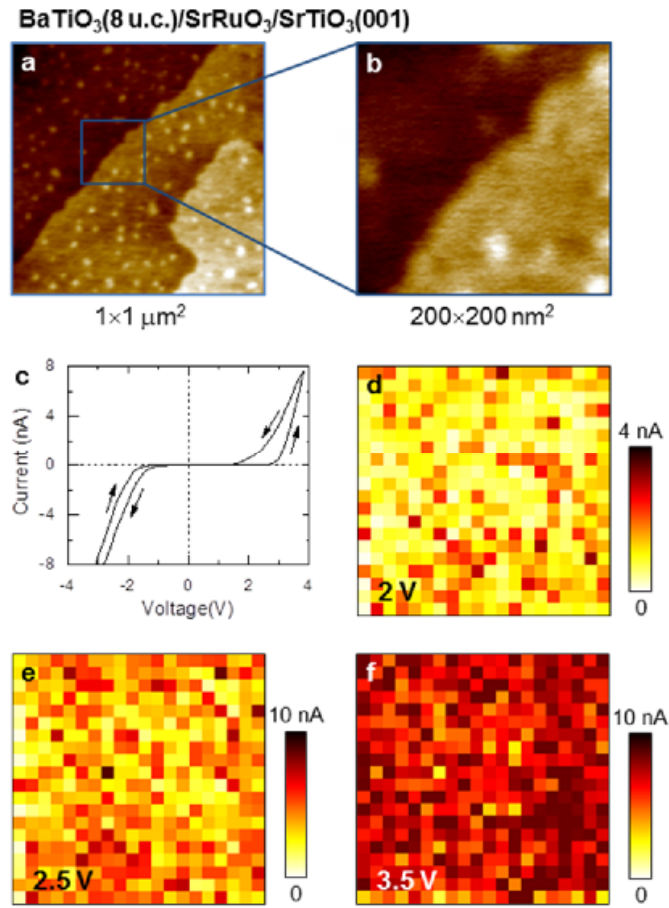

**Figure S18.** a,b) AFM topographic images of the 8 u.c. BTO film with a scanning area of (a) 1 × 1 μm² and (b) 200 × 200 nm². The bright dots in the images are one-unit-cell-high BTO islands due to imperfect thickness control. c) *I-V* curve averaged over the whole 20 × 20 pixels. d-f) Spatial maps of local current obtained at different tip biases. All the maps exhibit negligible conductance enhancement at the terrace edges.

## 10. Extended characterizations and discussions of FTJ fabrication

During the lift-off process for FTJ fabrication, the BTO film surface is inevitably exposed to various chemicals. These chemicals, especially water-based solutions, may cause contamination and formation of passive layers at the film surface.[26,27] This effect may lead to a big difference in the tunneling conductance measured through CAFM tips contact and FTJ. In order to prevent such surface contamination effects, we carefully chose to use organic solvent-based solutions only during the entire device fabrication process. Since the oxide samples were chemically inert to organic solvents, we expect that the surface contamination effect in our FTJ would be minimized.

We also performed two experiments to further exclude the surface contamination effect. First, we measured the AFM topographic image of BTO surface near an FTJ device. As shown in Figure S19a and S19b, the BTO film surface near the Ti/Au electrode still exhibits clear one-unit-cell-high terrace structure even after the lift-off process. Second, we directly compared the *I-V* curve measured through two geometries: 1) direct CAFM tip contact on the surface of as-grown BTO film; 2) FTJ device with Ti/Au top electrode. We choose the BTO film thickness as 6 u.c. to avoid any the terrace edge conductance. After normalizing the current according to contact area difference, the *I-V* curves measured in these two geometries (Figure S19c) exhibit similar shape and TER effect. These results indicate that the tunneling barrier is nearly unchanged after the FTJ fabrication procedure.

Based on the above results, we conclude that the lift-off process does not cause detectable contaminations to the BTO surface. In another word, except for the enhanced tunneling conductance at the terrace edges, the tunneling conductance through direct tip contact and FTJ device should not have any additional difference.

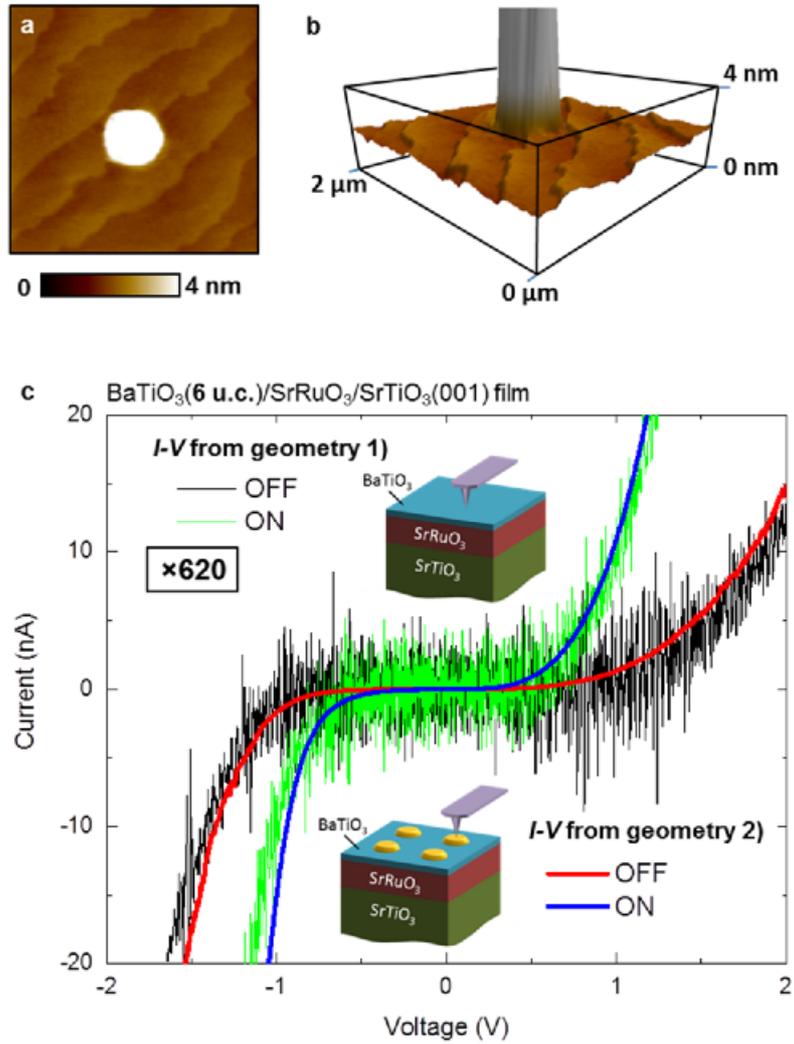

**Figure S19.** a,b) AFM topographic images of an FTJ device with an Au/Ti electrode [bright circle in (a) and high pillar in (b)]. After the lift-off process, the BTO film surface near the metal electrode still exhibits clear one-unit-cell-high terrace structure. c) $I_{ON}$-$V$ and $I_{OFF}$-$V$ curves measured from 6 u.c. BTO sample with two geometries: 1) direct CAFM tip contact on the surface of as-grown BTO film and 2) FTJ device with Ti/Au top electrode. Note that the $I$-$V$ curves measured in geometry 1) is multiplied by a factor of 620 to normalize the contact area difference with geometry 2). These curves are much noisier due to the small signal/noise ratio. The schematics of two measurement geometries are shown in the inset of (c).